\documentclass[aps,prd,floatfix,showpacs,superscriptaddress,nofootinbib,10pt]{revtex4-2}
\usepackage{amsmath}
\usepackage{amssymb}
\usepackage[Gray,squaren,thinqspace,thinspace]{SIunits}
\usepackage{subcaption}
\usepackage{graphicx}
\usepackage{braket}
\usepackage{bbm}
\usepackage{xcolor}
\newcommand{\mx}{\ensuremath{\mathsf}}
\DeclareMathOperator{\tr}{tr}
\DeclareMathOperator{\sgn}{sgn}
\newcommand{\MSbar}{\overline{\mbox{MS}}}

\begin{document}

\title{The gap equations of background field invariant Refined Gribov-Zwanziger action proposals and the deconfinement transition}
\author{David Dudal}\email{david.dudal@kuleuven.be}\affiliation{KU Leuven Campus Kortrijk -- Kulak, Department of Physics Etienne Sabbelaan 53 bus 7657, 8500 Kortrijk, Belgium}\affiliation{
	Ghent University, Department of Physics and Astronomy, Krijgslaan 281-S9, 9000 Gent, Belgium} \author{David Vercauteren}\email{vercauterendavid@duytan.edu.vn}\affiliation{Institute of Research and Development, Duy Tan University, Da Nang 550000, Vietnam}\affiliation{
	Faculty of Natural Sciences, Duy Tan University, Da Nang 550000, Vietnam}

\begin{abstract}
In earlier work, we set up an effective potential approach at zero temperature for the Gribov--Zwanziger model that takes into account not only the restriction to the first Gribov region as a way to deal with the gauge fixing ambiguity, but also the effect of dynamical dimension-two vacuum condensates. Here, we investigate the model at finite temperature in presence of a background gauge field that allows access to the Polyakov loop expectation value and the Yang--Mills (de)confinement phase structure. This necessitates paying attention to BRST and background gauge invariance of the whole construct. We employ two such methods as proposed elsewhere in literature: one based on using an appropriate dressed, BRST invariant, gluon field by the authors and one based on a Wilson-loop dressed Gribov--Zwanziger auxiliary field sector by Kroff and Reinosa. The latter approach outperforms the former, in estimating the critical temperature for $N=2,3$ as well as correctly predicting the order of the transition for both cases.
\end{abstract}

\maketitle

\section{Introduction}
It is well accepted from non-perturbative Monte Carlo lattice simulations that SU($N$) Yang--Mills gauge theories in the absence of fundamental matter fields undergo a deconfining phase transition at a certain critical temperature \cite{Lucini:2003zr,Lucini:2012gg}. This transition corresponds to the breaking of a global $\mathbb{Z}_N$ center symmetry when the Euclidean temporal direction is compactified on a circle, with circumference proportional to the inverse temperature \cite{Svetitsky:1985ye,Greensite:2003bk}. The vacuum expectation value of the Polyakov loop \cite{Polyakov:1978vu} serves as an order parameter for this symmetry, and has as such inspired an ongoing research activity into its dynamics, see for example \cite{Fukushima:2003fw,Schaefer:2007pw,Maas:2011ez,Fischer:2012vc,Dudal:2022nnu}.

Even in the presence of dynamical quark degrees of freedom (which explicitly break the center symmetry) the Polyakov loop remains the best observable to capture the cross-over transition, see \cite{Borsanyi:2010bp,Bazavov:2011nk} for ruling lattice QCD estimates. Since the transition temperature is of the order of the scale at which these gauge theories (which include QCD) become strongly coupled, it is a highly challenging endeavour to get reliable estimates for the Polyakov loop correlators, including its vacuum expectation value, analytically. This is further complicated by the non-local nature of the loop. These features highlight the sheer importance of lattice gauge theories to allow for a fully non-perturbative computational framework. Nonetheless, analytical takes are still desirable to offer a complementary view at the same physics, in particular as lattice simulations do also face difficulties when the physically relevant small quark mass limit must be taken, next to the issue of potentially catastrophic sign oscillations at finite density \cite{Fukushima:2010bq,deForcrand:2002hgr}.

Over the last two decades, a tremendous effort has been put into the development and application of Functional Methods to QCD, including the respective hierarchies of Dyson--Schwinger and Functional Renormalization Group equations \cite{vonSmekal:1997ohs, Alkofer:2000wg, Zwanziger:2001kw, Fischer:2003rp, Bloch:2003yu, Aguilar:2004sw, Boucaud:2006if, Aguilar:2007ie, Boucaud:2008ky, Fischer:2008uz, Rodriguez-Quintero:2010qad,Wetterich:1992yh, Berges:2000ew, Pawlowski:2003hq, Fischer:2004uk,Pawlowski:2005xe,Cyrol:2016tym,Dupuis:2020fhh,Maas:2011se} as well as variational approaches based on the Hamiltonian formulation or on $N$-particle-irreducible effective actions \cite{Schleifenbaum:2006bq,Quandt:2013wna,Quandt:2015aaa,Carrington:2007ea,Alkofer:2008tt,York:2012ib,Fister:2013bh} or alternatives \cite{Comitini:2017zfp}. These methods are quite successful in describing the vacuum properties of the theory as well as various aspects at finite temperature and/or density. They all rely, in one way or another, on the decoupling behavior of gluons in the Landau gauge, as dictated by results from lattice simulations \cite{Cucchieri:2007rg, Cucchieri:2008fc, Bornyakov:2008yx, Cucchieri:2009zt, Bogolubsky:2009dc, Bornyakov:2009ug, Dudal:2010tf,Duarte:2016iko}. More recently, a more phenomenological approach has been put forward based on the Curci--Ferrari model \cite{Tissier:2010ts,Tissier:2011ey,Reinosa:2015gxn,Reinosa:2016iml,Pelaez:2021tpq,Dudal:2022nnu}.

One particular way to deal with non-perturbative physics at the level of elementary degrees of freedom is by dealing with the Gribov issue \cite{Gribov:1977wm,Vandersickel:2012tz}: the fact that there is no unique way of selecting one representative configuration of a given gauge orbit in covariant gauges \cite{Singer:1978dk}. As there is also no rigorous way to deal properly with the existence of gauge copy modes in the path integral quantization procedure, in this paper we will use a well-tested formalism available to deal with the issue, which is known as the Gribov--Zwanziger (GZ) formalism: a restriction of the path integral to a smaller subdomain of gauge fields \cite{Gribov:1977wm,Zwanziger:1989mf,Zwanziger:1992qr}.

This approach was first proposed for the Landau and the Coulomb gauges . It long suffered from a serious drawback: its concrete implementation seemed to be inconsistent with BRST (Becchi--Rouet--Stora--Tyutin) invariance \cite{Becchi:1974xu,Becchi:1975nq,Tyutin:1975qk} of the gauge-fixed theory, which clouded its interpretation as a gauge (fixed) theory. Only more recently was it realized by some of us and colleagues how to overcome this complication to get a BRST-invariant restriction of the gauge path integral. As a bonus, the method also allowed the generalization of the Gribov--Zwanziger approach to the linear covariant gauges, amongst others \cite{Capri:2015ixa,Capri:2015nzw,Capri:2016aqq,Capri:2016gut}.

Another issue with the original Gribov--Zwanziger approach was that some of its major leading-order predictions did not match the corresponding lattice output. In the case of the Landau gauge, the Gribov--Zwanziger formalism by itself predicts, at tree level, a gluon propagator vanishing at momentum $p=0$, next to, more importantly, a ghost propagator with a stronger than $1/p^2$ singularity for $p\to0$. Although the latter fitted well in the Kugo--Ojima confinement criterion \cite{Kugo:1979gm}, it was at odds with large volume lattice simulations \cite{Cucchieri:2007md,Sternbeck:2007ug}. By now, several analytical takes exist on this, all compatible, qualitatively and/or quantitatively, with lattice data, not only for elementary propagators but also for vertices \cite{Dudal:2007cw,Dudal:2008sp,Aguilar:2008xm,Fischer:2008uz,Boucaud:2008ky,Binosi:2009qm,Tissier:2010ts,Gracey:2010cg,Tissier:2011ey,Bashir:2011dp,Maas:2011se,Dudal:2011gd,Boucaud:2011eh,Boucaud:2011ug,Vandersickel:2012tz,Aguilar:2012rz,Cucchieri:2012cb,Ayala:2012pb,Bashir:2012fs,Huber:2012kd,Serreau:2012cg,Rojas:2013tza,Pelaez:2013cpa,Eichmann:2014xya,Aguilar:2014lha,Reinosa:2014zta,Cyrol:2014kca,Blum:2014gna,Huber:2015ria,Capri:2015ixa,Capri:2015nzw,Siringo:2015gia,Aguilar:2015nqa,Binosi:2016wcx,Aguilar:2016vin,Eichmann:2016yit,Capri:2016aqq,Capri:2016aif,Capri:2016gut,Cyrol:2016tym,Pereira:2016inn,Reinosa:2017qtf,Capri:2017abz,Mintz:2017qri,Bermudez:2017bpx,Binosi:2017rwj,Cyrol:2017ewj,Comitini:2017zfp,Mintz:2018hhx,Huber:2018ned,Siringo:2018uho}.

In the Gribov--Zwanziger formalism in particular, the situation can be remedied by incorporating the effects of certain mass dimension-two condensates, the importance of which was already stressed before in papers like \cite{Gubarev:2000eu,Gubarev:2000nz,Boucaud:2001st,Verschelde:2001ia,Kondo:2001nq}. For the Gribov--Zwanziger formalism, this idea was first put on the table in \cite{Dudal:2007cw,Dudal:2008sp} with the condensate $\langle\bar\varphi\varphi-\bar\omega\omega\rangle$ (the fields here are Gribov localizing ghosts, see section \ref{GZ}). Later, a self-consistent computational scheme was constructed in \cite{Dudal:2011gd} based on the effective action formalism for local composite operators developed in \cite{Verschelde:1995jj,Verschelde:2001ia}, the renormalization of which was proven in \cite{Dudal:2002pq}. This construction is more natural with condensates like $\langle\bar\varphi\varphi\rangle$, $\langle\bar\varphi\bar\varphi\rangle$, and $\langle\varphi\varphi\rangle$. As the most promising candidate for a full description of the vacuum in this so-called ``refined Gribov--Zwanziger'' (RGZ) approach, the condensate $\langle\bar\varphi\varphi\rangle$ was considered in \cite{Dudal:2019ing} at zero temperature; this paper was meant as a jumping board for the present one. In the present work we consider both this last condensate and $\langle\bar\varphi\varphi-\bar\omega\omega\rangle$.

In \cite{Dudal:2017jfw}, the authors found that introducing a gluon background field into the Gribov--Zwanziger formalism (which is necessary to compute the vacuum expectation value of the Polyakov loop) is not as straightforward as one may naively be led to believe. A correct formalism was proposed in \cite{Dudal:2017jfw}, with a competing formalism later proposed by Kroff and Reinosa in \cite{Kroff:2018ncl}. In the present work, we again consider both these formalisms.

The structure of the paper is as follows. In Section II, we briefly sketch the original Gribov--Zwanziger approach at zero temperature in the Landau gauge, followed by a short reminder how to make this BRST invariant in Section III. Section IV deals with adding an appropriate background gauge to couple the Polyakov loop to the model and we summarize several approaches to do this in a BRST and background invariant fashion. In Section V, the addition of the dimension-two condensates is done, followed by preparatory computations at zero temperature in Section VI, needed to come to our finite temperature predictions in Section VII. We end with conclusions in Section VIII. Several technical results are relegated to a series of Appendices, including a constructive proof of a statement made in \cite{Kroff:2018ncl}. 

\section{A brief overview of the Gribov--Zwanziger formalism} \label{GZ}

Let us start by giving a short overview of the Gribov--Zwanziger framework
\cite{Gribov:1977wm,Zwanziger:1988jt,Zwanziger:1989mf,Zwanziger:1992qr}. As
already mentioned in the Introduction, the basic Gribov--Zwanziger action arises from the restriction of the domain of integration in the Euclidean
functional integral to the Gribov region $\Omega $, which is
defined as the set of all gauge field configurations fulfilling the Landau
gauge, $\partial _{\mu }A_{\mu }^{a}=0$, and for which the Faddeev--Popov
operator $\mathcal{M}^{ab}=-\partial _{\mu }(\partial _{\mu }\delta
^{ab}-gf^{abc}A_{\mu }^{c})$ is strictly positive, namely
\begin{equation*}
\Omega \;=\;\{A_{\mu }^{a}\;;\;\;\partial _{\mu }A_{\mu }^{a}=0\;;\;\;%
\mathcal{M}^{ab}=-\partial _{\mu }(\partial _{\mu }\delta
^{ab}-gf^{abc}A_{\mu }^{c})\;>0\;\}\;.
\end{equation*}%
The boundary $\partial\Omega$ of the region $\Omega$ is the (first) Gribov horizon.

One starts with the Faddeev--Popov action in the Landau gauge
\begin{subequations} \begin{equation} \label{fp}
S_{\text{FP}} = S_{\text{YM}} + S_{\text{Lg}} \;,
\end{equation}
where $S_{\text{YM}}$ and $S_{\text{Lg}}$ denote, respectively, the Yang--Mills and the Landau gauge-fixing terms, namely
\begin{gather}
	S_{\text{YM}} =\frac14 \int d^dx\;F_{\mu\nu}^aF_{\mu\nu}^a \;, \label{YM} \\
	S_{\text{Lg}} =\int d^dx\left( b^a\partial_\mu A_\mu^a+\bar c^a\partial_\mu D_\mu^{ab}c^b\right) \;, \label{gf}
\end{gather}
where $(\bar c^a,c^a)$ are the Faddeev--Popov ghosts, $b^{a}$ is the Lagrange multiplier implementing the Landau gauge, $D_{\mu}^{ab}=(\delta ^{ab}\partial _{\mu }-gf^{abc}A_{\mu }^{c})$ is the covariant derivative in the adjoint representation of $SU(N)$, and $F_{\mu \nu }^{a}$ denotes the field strength:
\begin{equation}
F_{\mu \nu }^{a}=\partial _{\mu }A_{\nu }^{a}-\partial _{\nu }A_{\mu
}^{a}+gf^{abc}A_{\mu }^{b}A_{\nu }^{c}\;. \label{fstr}
\end{equation} \end{subequations}
Following \cite{Gribov:1977wm,Zwanziger:1988jt,Zwanziger:1989mf,Zwanziger:1992qr}, the restriction of the domain of integration in the path integral is achieved by adding an additional term $H(A)$, called the horizon term, to the Faddeev--Popov action $S_{\text{FP}}$. This $H(A)$ is given by the following non-local expression
\begin{equation}
H(A,\gamma )={g^{2}}\int d^dxd^dy\ f^{abc}A_{\mu }^{b}(x)\left[
\mathcal{M}^{-1}(\gamma )\right] ^{ad}(x,y)f^{dec}A_{\mu }^{e}(y)\;,
\label{hf1}
\end{equation}%
where $\mathcal{M}^{-1}$ stands for the inverse of the Faddeev--Popov
operator. The partition function can then be written as \cite{Gribov:1977wm,Zwanziger:1988jt,Zwanziger:1989mf,Zwanziger:1992qr}:
\begin{equation} \label{zww1}
Z_{\text{GZ}} = \int_\Omega [\mathcal{D}A\mathcal{D}c\mathcal{D}\bar{c} \mathcal{D}b] e^{-S_{\text{FP}}} = \int [\mathcal{D}A\mathcal{D}c\mathcal{D}\bar{c}\mathcal{D}b] e^{-(S_{\text{FP}}+\gamma^4H(A,\gamma )-dV\gamma^4(N^{2}-1))} \;,
\end{equation}
where $V$ is the Euclidean space-time volume. The parameter $\gamma$ has the dimension of a mass and is known as the Gribov parameter. It is not a
free parameter of the theory. It is a dynamical quantity, being determined
in a self-consistent way through a gap equation called the horizon condition
\cite{Gribov:1977wm,Zwanziger:1988jt,Zwanziger:1989mf,Zwanziger:1992qr},
given by
\begin{equation} \label{hc1}
\langle H(A,\gamma)\rangle_{\text{GZ}} = dV(N^2-1) \;,
\end{equation}
where the notation $\langle \cdots \rangle_{\text{GZ}}$ means that the vacuum expectation value is to be evaluated with the measure defined in equation \eqref{zww1}. An equivalent all-order proof of equation \eqref{hc1} can be given within the original Gribov no-pole condition framework \cite{Gribov:1977wm}, by looking at the exact ghost propagator in an external gauge field \cite{Capri:2012wx}.

Although the horizon term $H(A,\gamma )$ in equation \eqref{hf1} is non-local, it can be cast in local form by means of the introduction of a set of auxiliary
fields $(\bar{\omega}_{\mu }^{ab},\omega _{\mu }^{ab},\bar{\varphi}_{\mu
}^{ab},\varphi _{\mu }^{ab})$, where $(\bar{\varphi}_{\mu }^{ab},\varphi
_{\mu }^{ab})$ are a pair of bosonic fields and $(\bar{\omega}_{\mu
}^{ab},\omega _{\mu }^{ab})$ are anti-commuting. It is not difficult to
show that the partition function $Z_{\text{GZ}}$ in eq.\eqref{zww1} can be
rewritten as \cite{Zwanziger:1988jt,Zwanziger:1989mf,Zwanziger:1992qr}
\begin{equation} \label{lzww1}
	Z_{\text{GZ}} = \int [\mathcal{D}\Phi] e^{-S_{\text{GZ}}[\Phi]} \;,
\end{equation}
where $\Phi$ accounts for the quantizing fields, $A$, $\bar{c}$, $c$, $b$, $\bar{\omega}$, $\omega $, $\bar{\varphi}$, and $\varphi $, while $S_{\text{GZ}}[\Phi]$ is the Yang--Mills action plus gauge fixing and Gribov--Zwanziger terms, in its localized version,
\begin{subequations} \begin{equation} \label{sgz}
	S_{\text{GZ}} = S_{\text{YM}} +S_{\text{gf}} +S_0 +S_\gamma \;,
\end{equation}
with
\begin{gather}
    S_0 = \int d^dx (\bar\varphi_\mu^{ac}(-\partial_\nu D_\nu^{ab})\varphi_\mu^{bc}-\bar\omega_\mu^{ac}(-\partial_\nu D_\nu^{ab})\omega_\mu^{bc}) \;, \label{s0} \\
    S_\gamma = \gamma^2g \int d^dx\ f^{abc}A_\mu^a (\varphi_\mu^{bc}+\bar\varphi_\mu^{bc}) - d\gamma^4 V (N^2-1) \;. \label{hfl}
\end{gather} \end{subequations}
It can be seen from \eqref{zww1} that the horizon condition \eqref{hc1}
takes the simpler form
\begin{equation} \label{ggap}
\frac{\partial \mathcal{E}_{v}}{\partial \gamma ^{2}}=0\;,
\end{equation}
which is called the gap equation. The quantity $\mathcal{E}_{v}(\gamma)$ is the vacuum energy defined by
\begin{equation}
	e^{-V\mathcal{E}_{v}}=Z_\text{GZ}\; \label{vce} \;.
\end{equation}

The local action $S_{\text{GZ}}$ in equation \eqref{sgz} is known as the Gribov--Zwanziger action. It has been shown to be renormalizable to all orders \cite{Zwanziger:1988jt,Zwanziger:1989mf,Zwanziger:1992qr,Maggiore:1993wq,Dudal:2007cw,Dudal:2008sp,Dudal:2010fq,Dudal:2011gd}. There are several issues with this action, though:
\begin{itemize}
	\item Its BRST invariance is softly broken. This has found a solution in \cite{Capri:2016aqq} through the $A^h$ formalism; this is reviewed in section \ref{sec:BRST_gluon}.
	\item The propagators of both gluons and ghosts are not in agreement with the lattice. This is remedied in the refined Gribov--Zwanziger (RGZ) formalism, which adds local composite operators (LCOs). This is reviewed in section \ref{LCO}.
\end{itemize}

\section{BRST-invariant gluon field $A^h$}\label{sec:BRST_gluon}
For a BRST-invariant formalism, it turns out to be most straightforward to introduce BRST-invariant projections of the gluon fields. This section gives a quick overview of the construction, which will be generalized in the following sections.

We start from the Yang--Mills action in a linear covariant gauge and in $d$ Euclidean space dimensions:
\begin{subequations} \begin{equation}\label{eq:LC}
	S_\text{LC} = S_\text{YM} + S_\alpha
\end{equation}
where $S_\alpha$ is now the gauge-fixing term in the linear covariant gauges:
\begin{equation}
	S_\alpha = \int d^dx (\tfrac\alpha2 b^a b^a + ib^a \partial_\mu A_\mu^a + \bar c^a \partial_\mu D_\mu^{ab}c^b) \;,
\end{equation} \end{subequations}
with $\alpha$ the gauge parameter. As we are eventually interested in imposing the Gribov restriction and introducing the dimension two gluon condensate $\langle A_\mu^2\rangle$ while preserving BRST invariance, we need a BRST invariant version of the $A_\mu^a$ field. In order to construct this, we insert the following unity into the path integral \cite{Capri:2018ijg,prep}:
\begin{subequations}
\begin{gather}
	1 = \mathcal N \int [\mathcal D\xi\mathcal D\tau\mathcal D\bar\eta\mathcal D\eta] e^{-S_h} \;, \label{unity} \\
	S_h = \int d^dx \left(i \tau^a\partial_\mu(A^h)_\mu^a + \bar\eta^a \partial_\mu (D^h)_\mu^{ab}\eta^b\right) \label{Sh} \;,
\end{gather}
where $\mathcal N$ is a normalization and $(D^h)_\mu^{ab}$ is the covariant derivative containing only the composite field $(A^h)_\mu^a$. This local but non-polynomial composite field object is defined as:
\begin{gather}
	(A^h)_\mu = h^\dagger A_\mu h + \tfrac ig h^\dagger \partial_\mu h \;, \label{3} \\
	h = e^{ig\xi} = e^{ig\xi^a T^a} \;,
\end{gather}
\end{subequations}
where the $T^a$ are the generators of the gauge group SU($N$). The $\xi^a$ are similar to Stueckelberg fields, while $\eta^a$ and $\bar\eta^a$ are additional (Grassmannian) ghost and anti-ghost fields. They serve to account for the Jacobian arising from the functional integration over $\tau^a$ to give a Dirac delta functional of the type $\delta(\partial_\mu(A^h)_\mu^a)$. That Jacobian is similar to the one of the Faddeev--Popov operator, and is supposed to be positive which amounts to removing a large class of infinitesimal Gribov copies, see \cite{Capri:2015ixa}. In mere perturbation theory, this is not the case, but the restriction to the Gribov region to be discussed will be sufficient to ensure it dynamically \cite{Gribov:1977wm,Zwanziger:1989mf}.

Expanding \eqref{3}, one finds an infinite series of local terms:
\begin{equation}\label{reeks}
	(A^h)_\mu^a = A_\mu^a - \partial_\mu\xi^a - gf^{abc}A_\mu^b\xi^c - \tfrac g2 f^{abc}\xi^b\partial_\mu\xi^c + \cdots \;.
\end{equation}
The unity \eqref{unity} can be used to stay within a local setup for an on-shell non-local quantity $(A^h)_\mu^a$ that can be added to the action. Notice that the multiplier $\tau^a$ implements $\partial_\mu(A^h)_\mu^a = 0$ which, when solved iteratively for $\xi^a$
\begin{subequations}
\begin{equation}
	\xi_* = \frac1{\partial^2} \partial_\mu A_\mu + ig \frac1{\partial^2} \left[\partial_\mu A_\mu,\frac1{\partial^2} \partial_\nu A_\nu\right] + \cdots \;,
\end{equation}
gives the (transversal) on-shell expression
\begin{gather}
	(A^h)_\mu = \left(\delta_{\mu\nu} - \frac{\partial_\mu\partial_\nu}{\partial^2} \right) \left(A_\nu + ig \left[A_\nu,\frac1{\partial^2} \partial_\lambda A_\lambda\right] + \cdots\right) \;,
\end{gather}
\end{subequations}
clearly showing the non-localities in terms of the inverse Laplacian. One can see that $A^h \to A$ when $A_\mu^a$ is in the Landau gauge $\partial_\mu A_\mu^a = 0$. We refer to \emph{e.g.} \cite{DellAntonio:1991mms,Lavelle:1995ty,Capri:2015ixa,Capri:2018ijg,Dudal:2022nnu,prep} for more details. It can be shown that $A^h$ is gauge invariant order per order, which is sufficient to establish BRST invariance. We will have nothing to say about large gauge transformations.

Mark that $(A^h)_\mu^a$ is formally the value of $A_\mu^a$ that (absolutely) minimizes the functional
\begin{equation}\label{A2}
	\int d^dx\ A_\mu^a A_\mu^a
\end{equation}
under (infinitesimal) gauge transformations $\delta A_\mu^a = D_\mu^{ab} \omega^b$, see e.g.~\cite{DellAntonio:1991mms,Lavelle:1995ty,Capri:2015ixa}. As such,
\begin{equation}\label{nummer}
\int d^dx (A^h)_\mu^a (A^h)_\mu^a= \min_{{\mbox{\tiny{gauge orbit}}}} \int d^dx\ A_\mu^a A_\mu^a \;,
\end{equation}
In practice, we are only (locally) minimizing the functional via a power series expansion \eqref{reeks} coming from infinitesimal gauge variations around the original gauge field $A_\mu^a$, whereas the extremum being a minimum is accounted for if the Faddeev--Popov operator (second order variation that is) is positive. This is discussed in \cite{Capri:2015ixa}.

This field $A^h$ can be used to construct a BRST-invariant modification of the Gribov--Zwanziger formalism. To do so, one replaces $S_0$ in \eqref{s0} with
\begin{subequations} \begin{equation} \label{s0h}
	S_{0h} = \int d^dx ({\bar{\varphi}}_{\mu }^{ac}(-\partial _{\nu }(D^h)_{\nu}^{ab})\varphi _{\mu }^{bc}-{\bar{\omega}}_{\mu }^{ac}(-\partial _{\nu}(D^h)_\nu^{ab})\omega_\mu^{bc}) \;,
\end{equation}
where $D^h$ is the covariant derivative with $A^h$ instead of $A$, and one replaces $S_\gamma$ in \eqref{hfl} with
\begin{equation} \label{hflh}
	S_{\gamma h} = \gamma^2g \int d^dx\ f^{abc}(A^h)_\mu^a(\varphi_\mu^{bc}+\bar\varphi_\mu^{bc}) - d\gamma^4V(N^2-1) \;.
\end{equation} \end{subequations}
The action $S_{\text{GZ}h} = S_{\text{YM}} + S_\alpha + S_h + S_{0h} +S_{\gamma h}$ enjoys the following exact BRST invariance, $sS_{\text{GZ}h} = 0$ and $s^2=0$ \cite{Capri:2015ixa}:
\begin{equation} \label{brstgamma} \begin{aligned}
s A^{a}_{\mu} =& -D^{ab}_{\mu}c^{b} \;, & s c^{a} =& \frac{g}{2}f^{abc}c^{b}c^{c}\;,\\
s \bar{c}^{a} =& ib^{a}\;, & s b^{a} =& 0\;, \\
s \varphi^{ab}_{\mu} =& 0 \;, & s \omega^{ab}_{\mu} =& 0\;, \\
s\bar\omega^{ab}_{\mu} =& 0 \;, & s\bar\varphi^{ab}_{\mu} =& 0\;,\\
s\varepsilon^{a} =& 0 \;, & s (A^h)^{a}_{\mu} =& 0 \;, \\
s h^{ij} =& -ig c^a (T^a)^{ik} h^{kj}.
\end{aligned} \end{equation}

\section{Including the Polyakov loop}
Our aim is to investigate the confinement/deconfinement phase transition of Yang--Mills theory. The standard way to achieve this goal is by probing the Polyakov loop order parameter,
\begin{equation}
	\mathcal{P} = \frac{1}{N}\tr \Braket{P e^{ig\int_{0}^{\beta}dt \ A_{0}(t,x)}} \;,
\end{equation}
where $P$ denotes path ordering, needed in the non-Abelian case to ensure the gauge invariance of $\mathcal{P}$. In analytical studies of the phase transition involving the Polyakov loop, one usually imposes the so-called ``Polyakov gauge'' on the gauge field, in which case the time-component $A_{0}$ becomes diagonal and independent of (imaginary) time: $\langle A_{\mu}(x)\rangle = \langle A_{0}\rangle \delta_{\mu 0}$, with $\langle A_{0}\rangle$ belonging to the Cartan subalgebra of the gauge group. In the SU(2) case for instance, the Cartan subalgebra is one-dimensional and can be chosen to be generated by $T^3\equiv\sigma^3/2$, so that $\langle A^{a}_{0}\rangle = \delta^{a3}\langle A^3_0\rangle \equiv \delta^{a3} \langle A_0\rangle$. More details on Polyakov gauge can be found in \cite{Marhauser:2008fz,Fukushima:2003fw,Ratti:2005jh}. Besides the trivial simplification of the Polyakov loop, when imposing the Polyakov gauge it turns out that the quantity $\Braket{A_{0}}$ becomes a good alternative choice for the order parameter instead of $\mathcal{P}$, see \cite{Marhauser:2008fz} for an argument using Jensen's inequality for convex functions, see also \cite{Braun:2007bx,Reinhardt:2012qe,Reinhardt:2013iia}. For other arguments based on the use of Weyl chambers and within other gauges (see below), see \cite{Reinosa:2015gxn,Herbst:2015ona,Reinosa:2020mnx}.

As explained in \cite{Braun:2007bx,Marhauser:2008fz,Reinosa:2014ooa}, in the SU(2) case at leading order we then simply find, using the properties of the Pauli matrices,
\begin{equation}
	\mathcal{P}=\cos\frac{\langle r\rangle}2\;,
\end{equation}
where we defined
\begin{equation}
 r=g\beta A_0\;,
\end{equation}
with $\beta$ the inverse temperature. This way, $r=\pi$ corresponds to the ``unbroken symmetry phase'' (confined or disordered phase), equivalent to $\Braket{{\cal P}} = 0$; while $r\not=\pi$ (modulo $2\pi$) corresponds to the ``broken symmetry phase'' (deconfined or ordered phase), equivalent to $\Braket{{\cal P}} \neq 0$. Since $\mathcal{P}\propto e^{-F/T}$ with $T$ the temperature and $F$ the free energy of a heavy quark, it is clear that in the unbroken phase (where the center symmetry is manifest: $\Braket{{\cal P}} = 0$), an infinite amount of energy would be required to free a quark. The broken/restored symmetry referred to is the $\mathbb{Z}_N$ center symmetry of a pure gauge theory (no dynamical matter in the fundamental representation). With a slight abuse of language, we will refer to the quantity $r$ as the Polyakov loop hereafter.

It is however a highly non-trivial job to actually compute $r$. An interesting way around was worked out in \cite{Braun:2007bx,Marhauser:2008fz,Reinosa:2014ooa}, where it was shown that similar considerations apply in Landau--DeWitt gauges, a generalization of the Landau gauge in the presence of a background. The background needs to be seen as a field of gauge-fixing parameters and, as such, can be chosen at will \emph{a priori}. However, specific choices turn out to be computationally more tractable while allowing one to unveil more easily the center-symmetry breaking mechanism. For the particular choice of self-consistent backgrounds which are designed to coincide with the thermal gluon average at each temperature, it could be shown that the background becomes an order parameter for center-symmetry as it derives from a center-symmetric background effective potential. An important assumption for this procedure to work is the underlying BRST invariance of the action, see \cite{Reinosa:2014ooa,Dudal:2022nnu}). 

In the presence of a gluon background field, the total gluon field is split into the background and the quantum fluctuations. We use the notation
\begin{equation}
	a_\mu^a = \bar A_\mu^a + A_\mu^a \;,
\end{equation}
where $a_\mu^a$ is the full gluon field, $\bar A_\mu^a$ is the background (which will correspond to the Polyakov loop), and $A_\mu^a$ are the quantum fluctuations around the background. Furthermore will write $\bar D_\mu^{ab} = \delta^{ab} \partial_\mu - gf^{abc} \bar A_\mu^c$ for the covariant derivative using only the background field $\bar A$. The gauge is fixed by replacing $S_{\text{Lg}}$ in \eqref{gf} by
\begin{equation} \label{sldw}
	S_{\text{LdW}} = \int d^dx ( b^a\bar D_\mu^{ab} (\bar A^b_\mu+A^b_\mu) + \bar c^a \bar D^{ab}_\mu (\bar D^{bc}_\mu-gf^{bcd}A^d_\mu)c^c ) \;.
\end{equation}

Two ways to add a background field to the Gribov--Zwanziger formalism have appeared in the literature: one that introduces a gauge-invariant background field $(\bar A^h)_\mu^a$ \cite{Dudal:2017jfw,Justo:2022vwa}, and one that ensures background gauge invariance by introducing non-local Wilson lines in the action \cite{Kroff:2018ncl}. We give a short review of both approaches in the subsections below.

\subsection{$\bar A^h$ approach}
In the $\bar A^h$ approach, the action is $S_h = S_{\text{YM}} + S_{\text{LdW}} + S_{0\text{LdW}h} + S_{\gamma\text{LdW}h} + S_{\text{LdW}h}$ with
\begin{subequations} \begin{gather}
	S_{0\text{LdW}h} = \int d^dx (\bar\varphi_\mu^{ad} (\bar D^h)_\mu^{ab} (D^h)_\mu^{bc} \varphi_\mu^{cd} - \bar\omega_\mu^{ad} (\bar D^h)_\mu^{ab} (D^h)_\mu^{bc} \omega_\mu^{cd}) \;, \label{s0ldwh} \\
	S_{\gamma\text{LdW}h} = \gamma^2g \int d^dx\ f^{abc} [(a^h)_\mu^a -(\bar A^h)_\mu^a] (\varphi_\mu^{bc} + \bar\varphi_\mu^{bc}) - d V (N^2-1)\gamma^4 \;, \\
	S_{\text{LdW}h} = \int d^dx \left(i \tau^a (\bar D^h)_\mu^{ab} ((a^h)_\mu^b - (\bar A^h)_\mu^b) + \bar\eta^a (\bar D^h)_\mu^{ab} (D^h)_\mu^{bc}\eta^c\right) \;.
\end{gather} \end{subequations}
In these expressions, $a^h$ is a transversal projection of the gluon field, $(D^h)_\mu^{ab} = \delta^{ab} \partial_\mu - gf^{abc} (a^h)_\mu^c$ is the covariant derivative using this $a^h$ field, and $\bar D^h$ is the covariant derivative containing $\bar A^h$, the background in the minimal Landau gauge (\emph{i.e.}~in the absolute minimum of \eqref{abarkw} \footnote{Mark that any $\bar A_\mu^a = \delta_{\mu0} \delta^{ai} rT/g$ for $i$ in the Casimir obeys the Landau gauge $\partial_\mu \bar A_\mu^a = 0$, but this is not the \emph{minimal} Landau gauge aimed for.}). Notice that, when coupling the gauge transformed gauge field $a^h$ to the localizing auxiliary fields $(\bar\varphi,\varphi)$, we used $a^h - \bar A^h$. This is because we are only interested in imposing the Gribov condition on the quantum fields, which are the fields we integrate over. This way the series of $a^h - \bar A^h$ starts at first order in the quantum gauge fields. For the rationale hereof, see \cite{Dudal:2017jfw}. Furthermore, mark that this approach applies the Gribov construction to the operator $-(\bar D^h)_\mu^{ab} (D^h)_\mu^{bc}$. The proof that this is sufficient is analogous to the one given in \cite{Dudal:2017jfw} and is for our case worked out in \appendixname\ \ref{DDproof}.

Let us start with the background and put it in the minimal Landau gauge. This means we minimize
\begin{equation} \label{abarkw}
	\int d^dx\ \bar A_\mu^a \bar A_\mu^a
\end{equation}
over the gauge orbit. If (for SU(2)) we start from a constant $\bar A_0^3 = r T/g$, this means we need to bring $r$ to a value $-2\pi<r<2\pi$. The case for more that two colors is analogous.

The quantum fields are to be put in the Landau background gauge. To construct $(A^h)_\mu^a$, we will use the background in its minimal Landau gauge form $(\bar A^h)_\mu^a$, such that we will require $(\bar D^h)_\mu^{ab} (a_\mu^b - (\bar A^h)_\mu^b) = 0$. This can be obtained from minimization of
\begin{equation}\label{minin}
	\int d^dx \Big(a_\mu^a-(\bar A^h)_\mu^a\Big) \Big(a_\mu^a-(\bar A^h)_\mu^a\Big) \;.
\end{equation}
This corresponds to the recipe used in \cite{Dudal:2017jfw}, with the important remark that for this paper we still worked at $T=0$ with constant background fields $\bar A^h$ in mind, effectively leading to $\bar A^h=0$. At $T>0$ and for the type of background gauge fields that interests us here, this is no longer true. 

In \cite{Justo:2022vwa}, the case was made to keep working with $a^h$ coming from minimizing $\int a^2$, as this leads to both BRST and background gauge invariance of the Gribov--Zwanziger action. This is true, but a price is paid: the classical (background) sector enters the Gribov construction, not only the quantum fields. It is not yet clear how the approach outlined in \cite{Justo:2022vwa} would deal with the terms that are linear in the quantum fields and which will enter the effective action due to this setup. We will therefore not consider the framework of \cite{Justo:2022vwa} for what follows.

To minimize \eqref{minin}, let us work in a series in the quantum field. Starting from $a_\mu^a$ we can perform a gauge transform
\begin{equation}
	a_\mu \to h^\dagger a_\mu h + \frac ig h^\dagger\partial_\mu h \;,
\end{equation}
where $a_\mu = a_\mu^a\tau^a/2$. Expand the matrix of the gauge transform as $h=h_0+h_1+\cdots$, where $h_0$ is the gauge transform matrix bringing $\bar A_\mu^a$ to $(\bar A^h)_\mu^a$, $h_1$ is first order in the quantum fields, and so on. Going to first order in the quantum fields, we have that
\begin{equation}
	a_\mu^h - \bar A_\mu^h = h_0^\dagger A_\mu h_0 + \frac ig \bar D_\mu^h (h_0^\dagger h_1) + \cdots \;.
\end{equation}
Applying the gauge condition yields
\begin{equation}
	\frac ig h_0^\dagger h_1 = - \frac1{\bar D^2_h} \bar D_\mu^h (h_0^\dagger A_\mu h_0) + \cdots\;,
\end{equation}
and some more algebra gives
\begin{equation} \label{ahfinal}
	a_\mu^h - \bar A_\mu^h = \left(\delta_{\mu\nu} - \bar D^h_\mu \frac1{\bar D^2_h} \bar D^h_\nu\right) (h_0^\dagger A_\nu h_0) + \cdots \;.
\end{equation}
We thus see that $a^h$ is attained by first gauge transforming $A_\mu^a$ using the adjoint of the gauge transform that set the background $\bar A_\mu^a$ equal to its lowest value, after which a certain projection operator must be applied.

Let us now look at what the result \eqref{ahfinal} entails for the physics of the theory. % After integrating out the Nakanishi--Lautrup field $b$, we find that the part of the action containing the gluon quantum fields up to quadratic terms is given by
%\begin{equation}\label{dv1}
%	\int d^dx \left( A_\mu^a \left( - \delta_{\mu\nu} (\bar D^2)^{ac} + \left(1-\tfrac1\alpha\right) \bar D_\mu^{ab}\bar D_\nu^{bc} \right) A_\nu^c + \gamma^2g f^{abc} [(a^h)_\mu^a -(\bar A^h)_\mu^a] (\varphi_\mu^{bc} + \bar\varphi_\mu^{bc}) \right) \;,
%\end{equation}
%where $\alpha$ is again the gauge parameter, which we will take to zero. Now we can replace the variable $A_\mu \to h_0^\dagger A_\mu h_0$ in the path integral. In the first term, the gauge matrices can be pulled through the background covariant derivatives, which will gauge transform the background field there: $\bar A_\mu \to \bar A_\mu^h$. In the second term, we simply get $a^h_\mu - \bar A^h_\mu \to \left(\delta_{\mu\lambda} - \bar D^h_\mu\frac1{\partial_\nu \bar D^h_\nu} \partial_\lambda \right) A_\lambda$.
%
%We conclude that the part of the action containing the gluon quantum fields up to quadratic terms becomes
We can always do a background gauge transformation on $\bar A_\mu$, $A_\mu$, $\bar c$, $c$, and $b$ using the gauge matrix $h_0$. This will have the effect that all background gauge fields $\bar A_\mu$ in the parts $S_{\text{YM}}$ and $S_{\text{LdW}}$ become $\bar A_\mu^h$; the parts $S_{0\text{LdW}h}$, $S_{\gamma\text{LdW}h}$, and $S_h$ remain unchanged as the gluon fields there appear in invariant combinations. % The factors of $h_0$ will drop out in \eqref{ahfinal}, though, such that we can write $S_{\gamma\text{LdW}h}$ as
%\begin{equation} \label{finalall}
%	S_{\gamma\text{LdW}h}' = \gamma^2g \int d^dx\ f^{abc} (\varphi_\mu^{bc} + \bar\varphi_\mu^{bc}) \left(\delta_{\mu\lambda}\delta^{ae} - \left(\bar D^h_\mu\frac1{\partial_\nu \bar D^h_\nu} \partial_\lambda \right)^{ae}\right) A_\lambda^e \;.
%\end{equation}
Finally, once we have imposed the Landau--DeWitt gauge through $S_{\text{LdW}}$ (see \eqref{sldw}), the projection operator in \eqref{ahfinal} will simplify to a unit operator and we have that $a_\mu^h - \bar A_\mu^h \to A_\mu + \cdots$.

It remains to discuss the BRST and background gauge invariance of \eqref{ahfinal}, order per order in the quantum fields. Intuitively, it is clear that we will find a BRST invariant $a^h$, since it corresponds to the minimum along the gauge orbit and BRST transformations correspond to local gauge transformations. To be more concrete, in the current case we have the following BRST symmetry generated by the operator $s$:\footnote{In \cite{brstbackground}, a nonzero transformation of the background gauge field $s\bar A_\mu=\Omega_\mu$ with $\Omega_\mu$ an auxiliary background ghost field was used, but this is not necessary for our purposes here. It merely served to simplify the algebraic discussion and proof of renormalizability of \cite{brstbackground}. The physical case is recovered when $\Omega_\mu^a\to0$, such that $(a^h)_\mu^a$ is invariant.}
\begin{equation}
 s\bar A_\mu^a = 0 \;, \qquad sA_\mu^a = -D_\mu^{ab} c^b \;, \qquad sc^a = \tfrac12 gf^{abc}c^bc^c \;, \qquad s\bar c^a = -ib^a \;,
\end{equation}
and all other transformations zero. This transformation gives, to leading order in the quantum fields
\begin{equation}
 s(h_0^\dagger A_\mu h^0) = -h_0^\dagger (D_\mu c) h_0 = -h_0^\dagger (\bar D_\mu c) h_0 + \cdots = - \bar D_\mu^h (h_0^\dagger ch_0) \;,
\end{equation}
such that \eqref{ahfinal} is indeed invariant.

Showing background gauge invariance is straightforward: transforming the background with some adjoint matrix $U$ needs to be undone by $h_0 \to U^\dagger h_0$ so as to keep $(\bar A^h)_\mu^a$ at its minimal value. This then requires a gauge transform with $U$ on $A_\mu^a$, $c^a$, $\bar c^a$, $b^a$, $\tau^a$, $\eta^a$, and $\bar\eta^a$ transforming as matter fields ($\Phi\to U^\dagger\Phi U$) while the Gribov ghosts $\varphi_\mu^{ab}$, $\bar\varphi_\mu^{ab}$, $\omega_\mu^{ab}$, and $\bar\omega_\mu^{ab}$ remain invariant. One easily verifies that this then leaves the action invariant.

\subsection{Kroff--Reinosa approach}
In the Kroff--Reinosa (KR) approach, the action is $S_h = S_{\text{YM}} + S_{\text{LdW}} + S_{0\text{KR}} + S_{\gamma\text{KR}}$ with
\begin{subequations} \begin{gather}\label{kr0}
	S_{0\text{KR}} = \int d^dx (\hat{\bar\varphi}_\mu^{ae} D_\nu^{ab} D_\nu^{bc} \hat\varphi_\mu^{ce} - \hat{\bar\omega}_\mu^{ae} D_\nu^{ab} D_\nu^{bc} \hat\omega_\mu^{ce}) \;, \\
	S_{\gamma\text{KR}} = \gamma^2g \int d^dx\ f^{abc} [a_\mu^a - \bar A_\mu^a] (\varphi_\mu^{bc} + \bar\varphi_\mu^{bc}) - d V (N^2-1)\gamma^4 \;,
\end{gather}
The hatted quantities here are defined as
\begin{equation} \label{WilsonKR}
	\hat\Phi_\mu^{ab}(x) = \Phi_\mu^{ac}(x) \left(P e^{ig\int_C dx'_\nu \bar A_\nu^e(x') T^e}\right)^{cb} \;,
\end{equation} \end{subequations}
for $\Phi$ equal to $\varphi$ or $\omega$, and the Hermitian adjoint hereof for $\bar\varphi$ and $\bar\omega$. The path $C$ connects the point $x$ to some arbitrary and constant point $x_0$, which (for the constant backgrounds we consider) does not influence the dynamics in any way \cite{Kroff:2018ncl}. Under gauge transformations of the background, the hatted quantities transform as matter fields with only \emph{one} index, as the path-ordered exponential in \eqref{WilsonKR} absorbs the background gauge transformation of the second index. This ensures the background invariance of the action.

In practice, the effect of the Wilson line in \eqref{WilsonKR} is rather technical to work out, but when the dust settles and one integrates out the $(\bar\varphi,\varphi)$ fields, one obtains the gluon propagator term
\begin{equation}
	2g^2(N^2-1)\gamma^4 \delta_{\mu\nu} \left(\frac1{-\bar D^2}\right)^{ab} \;,
\end{equation}
as was used in \cite{Canfora:2015yia}. The structure constants that usually flank the inverse Faddeev--Popov operator in this term are absent, which greatly simplifies the computations.

Kroff \& Reinosa also proposed to introduce color-dependent Gribov parameters:
\begin{equation} \label{cdGp}
	\Big(\gamma_0P^{ab} + \gamma_{\text{ch}} (\delta^{ab}-P^{ab})\Big) A_\mu^b \;,
\end{equation}
where $P^{ab}$ is a projection operator on the ``neutral'' subspace of color space (in the terminology of \cite{Kroff:2018ncl}), see \appendixname\ \ref{KRproj} for the explicit construction of this non-trivial operator, which we did not find in \cite{Kroff:2018ncl}. We will not consider the nondegenerate case, where there are $N^2-1$ different Gribov parameters, but only the partially degenerate case, where all the Gribov parameters in the ``charged'' subspace are taken equal and denoted $\gamma_{\text{ch}}$.

In \cite{Kroff:2018ncl}, the authors note the loss of BRST invariance. As we already stressed the importance of this BRST invariance to ensure that a physical (background) effective potential can be computed \cite{Reinosa:2014ooa,Dudal:2022nnu}, let us spend a few words here to show that the Kroff--Reinosa construction can be recast in a BRST-invariant formulation. On shell and in the Landau--DeWitt gauge, this will effectively collapse back to \eqref{kr0}, \emph{a posteriori} granting credit to the approach of \cite{Kroff:2018ncl}. The construction again relies on the definition of a BRST-invariant $A^h$ field. However, given that the Kroff--Reinosa setup is already manifestly invariant under gauge transformations of the background, the $h_0$ used in the previous subsection is spurious. (Remember that in the Kroff--Reinosa setup, the auxiliary fields transforms in the bi-adjoint. So using the construct \eqref{ahfinal} is not an option here, since it does not transform under background transformations.) This means we need an approach similar to the one used in \cite{Dudal:2022nnu}.

As such, we minimize
\begin{equation}
	\int d^dx (a_\mu^a-\bar A_\mu^a)^2 = \int d^dx (A_\mu^a)^2
\end{equation}
under infinitesimal gauge transformations $\delta a_\mu^a = \delta A_\mu^a = D_\mu^{ab} \omega^b$ to find a field $(A^h)_\mu^a$ (and the background does not transform, see \cite{Zwanziger:1982na,Cucchieri:2012ii} for more details). Then in $S_{0\text{KR}}$ we make the replacement $D D \to D^h D^h$, where $(D^h)_\mu^{ab}$ is the covariant derivative containing $\bar A_\mu^a + (A^h)_\mu^a$. This makes this part of the action BRST invariant.  The part $S_{\gamma\text{KR}}$ already transforms correctly.

\section{BRST-invariant condensates} \label{LCO}
This section presents a short review of the Local Composite Operator (LCO) formalism as proposed in \cite{Verschelde:2001ia} modified in the presence of a background field and the Gribov horizon.

\subsection{Dimension-two gluon condensate}
A BRST analysis \cite{Dudal:2022nnu} (for BRST in the background gauge, see for example \cite{Ferrari:2000yp,brstbackground}) shows that, for the LCO formalism to stay renormalizable, the dimension-two operator
\begin{equation}
(a_\mu^h - \bar A_\mu^h)^2
\end{equation}
should be used. First, the source terms
\begin{equation} \label{lcobg}
\int d^dx \left(\tfrac12 J (a_\mu^h - \bar A_\mu^h)^2 - \tfrac12 \zeta J^2\right)
\end{equation}
are added to the action with $J$ the source used to couple the operator to the theory. The term in $J^2$ is necessary here for renormalizability of the generating functional of connected diagrams $W(J)$ and, subsequently, of the associated generating functional of 1PI diagrams $\Gamma$, known as the effective action. Here $\zeta$ is a new coupling constant whose determination we will discuss later. In the physical vacuum, corresponding to $J\to0$, it should decouple again, at least if we were to do the computations exactly. At (any) finite order, $\zeta$ will be explicitly present, even in physical observables, making it necessary to choose it as wisely as possibly. Notice that $\zeta$ is \emph{not} a gauge parameter as it in fact couples to the BRST invariant quantity $J^2$. Indeed, in a BRST invariant theory, we expect the gauge parameter to explicitly cancel order per order from physical observables, a fact guaranteed by \emph{e.g.}~the Nielsen identities \cite{Nielsen:1975fs}, which are in themselves a consequence of BRST invariance \cite{Piguet:1984js}. Thanks to $\zeta$, the Lagrangian remains multiplicatively renormalizable (see \cite{Dudal:2022nnu}).

To actually compute the effective potential, it is computationally simplest to rely on Jackiw's background field method \cite{Jackiw:1974cv}. Before integrating over any fluctuating quantum fields, a Legendre transform is performed, so that formally $\sigma=\frac12 (a_\mu^h - \bar A_\mu^h)^2-\zeta J$. Plugging this into the Legendre transformation between $\Gamma$ and $W$, we find that we could just as well have started from the original path integral with the following unity inserted into it:\footnote{We normalize $\sigma$ like in \cite{Dudal:2011gd}.}
\begin{equation}\label{uniteit2}
	1 = \mathcal N \int[\mathcal D\sigma] e^{-\frac12 \int d^dx \left(\sigma + \tfrac1{2\sqrt\zeta}(a_\mu^h - \bar A_\mu^h)^2\right)^2} \;,
\end{equation}
with $\mathcal N$ an irrelevant constant. This is equivalent to a Hubbard--Stratonovich transformation, see for instance \cite{Verschelde:2001ia,prep}, and it also evades the interpretational issues for the energy when higher-than-linear terms in the sources are present. Of course, if we could integrate the path integral exactly, this unity would not change a thing. The situation only gets interesting if the perturbative dynamics of the theory assign a non-vanishing vacuum expectation value to $\sigma$. As such, this $\sigma$ field allows to include potential non-perturbative information through its vacuum expectation value. In the case without a background, $\sigma$ does indeed condense and a vacuum with $\Braket\sigma\not=0$ is preferred.

For the record, BRST invariance is ensured if we assign {$s\sigma=-s\left(\tfrac12(a_\mu^h - \bar A_\mu^h)^2\right)$,} which implies off-shell that $s\sigma=0$ thanks to the BRST invariance of $a_\mu^h-\bar A_\mu^h$.

It is evident that $\zeta$ can be interpreted as a genuine new coupling constant. Therefore, we now have two coupling constants, $g^2$ and $\zeta$, with $g^2$ running as usual, that is: independently of $\zeta$. This makes our situation suitable for the Zimmermann reduction of couplings program \cite{Zimmermann:1984sx}, see also \cite{Heinemeyer:2019vbc} for a recent overview. In this program, one coupling ($\zeta$ in our case) is re-expressed as a series in the other (here $g^2$), so that the running of $\zeta$ controlled by $\zeta(g^2)$ is then automatically satisfied, see also \cite{prep}. More specifically, $\zeta(g^2)$ is determined such that the generating functional of connected Green functions, $W(J)$, obeys a standard, linear renormalization group equation \cite{Verschelde:2001ia}.

This selects one consistent coupling $\zeta(g^2)$ from a whole space of allowed couplings, and it is also the unique choice compatible with multiplicative renormalizability \cite{Verschelde:2001ia}. Given that $\zeta$ should, in principle, not affect physics, we can safely rely here on this special choice, already made earlier in \emph{e.g.} \cite{Verschelde:2001ia}. This choice seems also to be a natural one from the point of view of the loop expansion of the background potential to be used below. In the $\MSbar$ scheme, one finds \cite{Verschelde:2001ia,7}
\begin{subequations} \label{zetadeltazeta}
\begin{gather}
\zeta = \frac{N^2-1}{g^2N} \left(\frac{9}{13} + \frac{g^2N}{16\pi^2}\frac{161}{52} + \mathcal O(g^4)\right)\;, \\
Z_\zeta = 1 - \frac{g^2N}{16\pi^2} \frac{13}{3\epsilon} + \mathcal O(g^2)\;, \\
Z_J = 1 - \frac{Ng^2}{16\pi^2} \frac{35}{6\epsilon} + \mathcal O(g^2)\;,
\end{gather}
\end{subequations}
where $Z_\zeta$, $Z_J$ are the renormalization factors of $\zeta J^2$, $J$ respectively.

\subsection{Refined Gribov--Zwanziger action} \label{RGZ formalism}
In \cite{Dudal:2007cw,Dudal:2008sp,Dudal:2011gd}, it was noticed that the Gribov--Zwanziger formalism in Landau gauge is disturbed by non-perturbative dynamical instabilities, caused by the formation of dimension-two condensates, $\langle A_{\mu}^{a}A_{\mu}^{a}\rangle$, $\langle \bar{\varphi}^{ab}_{\mu}\varphi^{ab}_{\mu}-\bar{\omega}^{ab}_{\mu}\omega^{ab}_{\mu} \rangle$, and/or $\langle\bar{\varphi}^{ab}_{\mu}\varphi^{ab}_{\mu}\rangle$, which are energetically favored. Similar features were later noticed in the Maximal Abelian gauge Gribov--Zwanziger formulation \cite{Capri:2010an,Capri:2017abz}. This led to the Refined Gribov--Zwanziger formalism, that explicitly takes the effects of these condensates into account.

The construction for the localizing-ghost condensates is analogous to that for the dimension-two gluon condensate. For the couplings and renormalization factors involved, we refer to the literature, see e.g.~\cite{Verschelde:2001ia,Dudal:2019ing,Dudal:2022nnu} and references therein.

The original proposal for the refinement to the Gribov--Zwanziger formalism \cite{Dudal:2008sp} used the symmetric condensate $\langle \bar\varphi^{ab}_\mu\varphi^{ab}_\mu-\bar\omega^{ab}_\mu\omega^{ab}_\mu \rangle$. This condensate has the advantage that it is immediately finite and strictly speaking no source-squared term (in the vein of the last term of \eqref{lcobg}) is necessary. As a result, however, the gap equation for the condensate has no nonperturbative solutions. The Hubbard--Stratonovich transformation becomes useless here and as a result, there is no ``classical'' quadratic part for the potential for the condensate. We will circumvent this issue in the following section.

 Using instead the philosophy of the approach starting from the analogon of \eqref{uniteit2} does not run into this problem, though, at the cost of introducing one truly new free coupling. In the following, we will call this approach the ``symmetric'' case.

Later approaches \cite{Dudal:2011gd} focused on the condensate $\langle\bar{\varphi}^{ab}_{\mu}\varphi^{ab}_{\mu}\rangle$. The $T=0$ case was fully explored in \cite{Dudal:2019ing}, which can be immediately used as the starting point for the study of the Polyakov loop. In the following, this approach will be called the ``$\bar\varphi\varphi$'' case.

\section{Zero temperature jumping board}
\subsection{Relevant parts of action}
To compute the effective action at first order in the quantum corrections, we need the background part (classical part) and the quadratic terms of the action (of which we will need the trace-logarithm to compute the first-order quantum corrections).

The first term of the semi-classical perturbation series consists of the background terms. We only consider backgrounds with $F_{\mu\nu}^a=0$, such that these background terms will only come from the LCO parts and from the Gribov--Zwanziger action. First we review some of the relevant formulae, which can be found in the literature.

From the Gribov--Zwanziger action we get, with the $Z$ factors restored and in the more general renormalization scheme of \cite{Dudal:2019ing},
\begin{equation}
	-dV(N^2-1)Z_\gamma^2\gamma^4 \;, \qquad Z_\gamma = 1 + \frac{b_0}2 \frac{Ng^2}{(4\pi)^2} + \frac38 \frac{Ng^2}{(4\pi)^2} \frac2\epsilon \;.
\end{equation}
In $d=4-\epsilon$, this gives
\begin{equation}
	-4V(N^2-1)\left(1 - \frac38 \frac{Ng^2}{(4\pi)^2} + b_0 \frac{Ng^2}{(4\pi)^2} + \frac34 \frac{Ng^2}{(4\pi)^2} \frac2\epsilon\right)\gamma^4 \;.
\end{equation}
To this we add the LCO part. The ``usual'' LCO part is:
\begin{equation}
	S_\text{LCO} = \int d^4x \left[ \frac12 \sigma^2 + \frac1{2\sqrt\zeta} \sigma (a_\mu^h - \bar A_\mu^h)^2 + \frac1{8\zeta} ((a_\mu^h - \bar A_\mu^h)^2)^2 \right] \;.
\end{equation}
To add the $\bar\varphi\varphi$ condensate, we need instead
\begin{align}
	S_{A^2+\bar\varphi\varphi} =& \int d^4x \left[ \frac12 \sigma_1^2 - \frac12 \sigma_2^2 + \frac1{2\sqrt\zeta} \sigma_1 (a_\mu^h - \bar A_\mu^h)^2 - \sqrt{\tfrac\zeta{-2\alpha\zeta+\chi^2}} \sigma_2 (\bar\varphi\varphi-\tfrac\chi{2\zeta} (a_\mu^h - \bar A_\mu^h)^2) \right. \nonumber \\ & \left. + \frac1{8\zeta} ((a_\mu^h - \bar A_\mu^h)^2)^2 - \frac12 \tfrac\zeta{-2\alpha\zeta+\chi^2} (\bar\varphi\varphi-\tfrac\chi{2\zeta} (a_\mu^h - \bar A_\mu^h)^2)^2 \right] \;.
\end{align}
The background part with the renormalization factors restored is just
\begin{equation}
	\int d^4x \left[ \frac1{2Z_\zeta} \sigma_1^2 - \frac1{2Z_\alpha} \sigma_2^2\right] \;, \qquad Z_\zeta^{-1} = 1 + \frac{13}6 \frac{Ng^2}{(4\pi)^2} \frac2\epsilon \;, \qquad Z_\alpha^{-1} = 1 - \frac{35}{12} \frac{Ng^2}{(4\pi)^2} \frac2\epsilon \;.
\end{equation}
In the symmetric case we can just generalize the normal LCO case (because there is no mixing):
\begin{align}
	S_\text{sym} =& \int d^4x \left[ \frac12 \sigma_1^2 + \frac1{2\sqrt\zeta} \sigma_1 (a_\mu^h - \bar A_\mu^h)^2 + \frac1{8\zeta} ((a_\mu^h - \bar A_\mu^h)^2)^2 \right] \nonumber \\ & + \int d^4x \left[ \frac12 \sigma_2^2 - \frac1{\sqrt\beta} \sigma_2 (\bar\varphi\varphi-\bar\omega\omega) + \frac1{2\beta} (\bar\varphi\varphi-\bar\omega\omega)^2 \right] \;.
\end{align}
Here, $\beta$ is a new (free) coupling constant that will require determination. This $\beta$ cannot be fixed from renormalization-group requirements as is the case with $\zeta$, this due to the aforementioned lack of quadratic divergences after introducing the symmetric condensate. This means $\beta$ is a non-running parameter that can be freely chosen; we determine a value for it in \appendixname\ \ref{paramsymm}.

The background part with the $Z$ factors restored is just
\begin{equation}
	\int d^4x \left[ \frac1{2Z_\zeta} \sigma_1^2 + \frac12 \sigma_2^2\right] \;, \qquad Z_\zeta^{-1} = 1 + \frac{13}6 \frac{Ng^2}{(4\pi)^2} \frac2\epsilon \;.
\end{equation}

We can now write down the background and quadratic parts for the cases we consider in this paper. At zero temperature, the gluon background field does not yet need to be included.

The full background part (classical part) in the $\bar\varphi\varphi$ case is
\begin{subequations} \begin{align}
	\int d^4x &\left[ -\frac{2(N^2-1)}{Ng^2} \left(1 - \frac38 \frac{Ng^2}{(4\pi)^2} + b_0 \frac{Ng^2}{(4\pi)^2} + \frac34 \frac{Ng^2}{(4\pi)^2} \frac2\epsilon\right)\lambda^4 \right. \nonumber \\ &\left. + \frac9{26} \frac{N^2-1}{Ng^2} \left(1 + \frac{13}6 \frac{Ng^2}{(4\pi)^2} \frac2\epsilon\right) m^4 - \frac{24}{35} \frac{(N^2-1)^2}{Ng^2} \left(1 - \frac{35}{12} \frac{Ng^2}{(4\pi)^2} \frac2\epsilon\right) M^4\right] \;.
\end{align}
where we defined
\begin{gather}
	\lambda^4 = 2Ng^2\gamma^4 \;, \\
	m^2 = \left.\frac1{\sqrt\zeta}\right|_\text{leading} \sigma_1 = \sqrt{\frac{13}9 \frac{Ng^2}{N^2-1}} \sigma_1 \;, \qquad M^2 = \left.\sqrt{\tfrac\zeta{-2\alpha\zeta+\chi^2}}\right|_\text{leading} \sigma_2 = \sqrt{\frac{35}{48} \frac{Ng^2}{(N^2-1)^2}} \sigma_2 \;.
\end{gather} \end{subequations}
The quadratic part of the action is:
\begin{subequations} \begin{multline}
	\int d^dx \left( \frac12 A_\mu^a \left( - \delta_{\mu\nu} \partial^2 + \left(1-\frac1\xi\right) \partial_\mu\partial_\nu \right) A_\nu^a + \bar c^a \partial^2 c^a + U_\mu^{ab} \partial^2 U_\mu^{ab} + V_\mu^{ab} \partial^2 V_\mu^{ab} \right. \\
	\left. - \bar\omega_\mu^{ab} \partial^2 \omega_\mu^{ab} - 2\gamma^2 g f^{abc} A_\mu^a U_\mu^{bc} + \frac{m^2}2 A^2 - M^2 (U^2+V^2) \right) \;.
\end{multline}
where
\begin{equation}
	U_\mu^{ab} = \frac12 (\varphi^{ab}_{\mu}+\bar{\varphi}_\mu^{ab}) \;, \qquad V_\mu^{ab} = \frac i2 (\varphi^{ab}_{\mu}-\bar{\varphi}_\mu^{ab}) \;.
\end{equation} \end{subequations}

The full background part (classical part) in the symmetric case is
\begin{subequations} \begin{equation}
	\int d^4x \left[ -\frac{2(N^2-1)}{Ng^2} \left(1 - \frac38 \frac{Ng^2}{(4\pi)^2} + b_0 \frac{Ng^2}{(4\pi)^2} + \frac34 \frac{Ng^2}{(4\pi)^2} \frac2\epsilon\right)\lambda^4 + \frac9{26} \frac{N^2-1}{Ng^2} \left(1 + \frac{13}6 \frac{Ng^2}{(4\pi)^2} \frac2\epsilon\right) m^4 + \frac\beta2 M^4\right] \;,
\end{equation}
where
\begin{equation}
	\lambda^4 = 2Ng^2\gamma^4 \;, \qquad m^2 = \left.\frac1{\sqrt\zeta}\right|_\text{leading} \sigma_1 = \sqrt{\frac{13}9 \frac{Ng^2}{N^2-1}} \sigma_1 \;, \qquad M^2 = \frac1{\sqrt\beta} \sigma_2 \;.
\end{equation} \end{subequations}
The quadratic part of the action is:
\begin{subequations} \begin{multline}
	\int d^dx \left( \frac12 A_\mu^a \left( - \delta_{\mu\nu} \partial^2 + \left(1-\frac1\xi\right) \partial_\mu\partial_\nu \right) A_\nu^a + \bar c^a \partial^2 c^a + U_\mu^{ab} \partial^2 U_\mu^{ab} + V_\mu^{ab} \partial^2 V_\mu^{ab} \right. \\
	\left. - \bar\omega_\mu^{ab} \partial^2 \omega_\mu^{ab} - 2\gamma^2 g f^{abc} A_\mu^a U_\mu^{bc} + \frac{m^2}2 A^2 - M^2 (U^2+V^2 - \bar\omega\omega) \right) \;.
\end{multline}
where
\begin{equation}
	U_\mu^{ab} = \frac12 (\varphi^{ab}_{\mu}+\bar{\varphi}_\mu^{ab}) \;, \qquad V_\mu^{ab} = \frac i2 (\varphi^{ab}_{\mu}-\bar{\varphi}_\mu^{ab}) \;.
\end{equation} \end{subequations}

\subsection{Effective actions at zero temperature}
The logarithmic trace of the operators is
\begin{multline} \label{trlns}
	\frac12 \tr\ln \begin{pmatrix} \delta^{ab} \left( \delta_{\mu\nu} (p^2+m^2) - \left(1-\frac1\xi\right) p_\mu p_\nu \right) & -2\gamma^2 g f^{aef} \delta_{\mu\nu} \\ -2\gamma^2 g f^{bcd} \delta_{\mu\nu} & -2\delta^{ce}\delta^{df}\delta_{\mu\nu} (p^2+M^2) \end{pmatrix} \\ - (N^2-1) \tr\ln(p^2) + \frac d2(N^2-1)^2 \tr\ln(p^2+M^2) - d(N^2-1)^2 \tr\ln(p^2+sM^2) \\
	= \frac12 (N^2-1) (d-1) \tr\ln \left(p^2+m^2 + \frac{\lambda^4}{p^2+M^2}\right) - \frac12 (N^2-1) \tr\ln(p^2) + d(N^2-1)^2 \tr\ln\frac{p^2+M^2}{p^2+sM^2} \;,
\end{multline}
were we took the limit $\xi\to0$, and $s=0$ for the $\bar\varphi\varphi$ approach and $s=1$ for the symmetric approach. The first $\tr\ln$ can be rewritten as
\begin{equation}
	\frac12 (N^2-1)(d-1) \bigg(\tr\ln (p^2+z_+) + \tr\ln (p^2+z_-) - \tr\ln (p^2+M^2)\bigg) \;,
\end{equation}
where
\begin{equation}
	z_\pm = \frac12 \left(m^2+M^2 \pm i\sqrt{4\lambda^4-(m^2-M^2)^2}\right) \;.
\end{equation}
Computing the trace in $d=4-\epsilon$ dimensions gives
\begin{equation}
	-\frac3{4(4\pi)^2} (N^2-1) \left( (z_+^2+z_-^2-M^4)\left(\frac2\epsilon+\frac56\right) - z_+^2\ln\frac{z_+}{\bar\mu^2} - z_-^2\ln\frac{z_-}{\bar\mu^2} + M^4\ln\frac{M^2}{\bar\mu^2} \right) \;.
\end{equation}
Given
\begin{subequations} \begin{gather}
	z_+^2+z_-^2-M^4 = m^4 - 2\lambda^4 \;, \\
	z_+z_- = m^2M^2 + \lambda^4 \;, \\
	z_+^2-z_-^2 = i (m^2+M^2) \sqrt{4\lambda^4-(m^2-M^2)^2} \;, \\
	z_+^2\ln\frac{z_+}{\bar\mu^2} + z_-^2\ln\frac{z_-}{\bar\mu^2} = \frac{z_+^2+z_-^2}2 \ln\frac{z_+z_-}{\bar\mu^4} + \frac{z_+^2-z_-^2}2\ln\frac{z_+}{z_-} \;, \\
	\ln\frac{z_+}{z_-} = 2i\arctan\frac{\sqrt{4\lambda^4-(m^2-M^2)^2}}{m^2+M^2} \;,
\end{gather} \end{subequations}
we get for the trace:
\begin{multline}
	-\frac3{4(4\pi)^2} (N^2-1) \left( (m^4 - 2\lambda^4)\left(\frac2\epsilon+\frac56\right) - \frac12(m^4+M^4 - 2\lambda^4)\ln\frac{m^2M^2 + \lambda^4}{\bar\mu^4} \right. \\ \left. + (m^2+M^2) \sqrt{4\lambda^4-(m^2-M^2)^2}\arctan\frac{\sqrt{4\lambda^4-(m^2-M^2)^2}}{m^2+M^2} + M^4\ln\frac{M^2}{\bar\mu^2} \right) \\
	= -\frac3{4(4\pi)^2} (N^2-1) \left( (m^4 - 2\lambda^4)\left(\frac2\epsilon+\frac56 - \frac12\ln\frac{m^2M^2 + \lambda^4}{\bar\mu^4}\right) \right. \\ \left. + (m^2+M^2) \sqrt{4\lambda^4-(m^2-M^2)^2}\arctan\frac{\sqrt{4\lambda^4-(m^2-M^2)^2}}{m^2+M^2} - \frac12 M^4\ln\frac{m^2M^2 + \lambda^4}{M^4} \right) \;.
\end{multline}

The last $\tr\ln$ is
\begin{equation}
	d(N^2-1)^2 \tr\ln\frac{p^2+M^2}{p^2+sM^2} = (1-s)d(N^2-1)^2 \tr\ln(p^2+M^2) = -\frac{2M^4}{(4\pi)^2}(1-s)(N^2-1)^2 \left(\frac2\epsilon+1-\ln\frac{M^2}{\bar\mu^2}\right) \;.
\end{equation}

In the $\bar\varphi\varphi$ approach we have
\begin{multline}
	\Gamma_{\bar\varphi\varphi}(m^2,M^2,\lambda^4) = -\frac{2(N^2-1)}{Ng^2} \left(1 - \frac38 \frac{Ng^2}{(4\pi)^2} + b_0 \frac{Ng^2}{(4\pi)^2}\right)\lambda^4 + \frac9{26} \frac{N^2-1}{Ng^2} m^4 - \frac{24}{35} \frac{(N^2-1)^2}{Ng^2} M^4 \\
	-\frac3{4(4\pi)^2} (N^2-1) \left( (m^4 - 2\lambda^4)\left(\frac56 - \frac12\ln\frac{m^2M^2 + \lambda^4}{\bar\mu^4}\right) \right. \\ \left. + (m^2+M^2) \sqrt{4\lambda^4-(m^2-M^2)^2}\arctan\frac{\sqrt{4\lambda^4-(m^2-M^2)^2}}{m^2+M^2} - \frac12 M^4\ln\frac{m^2M^2 + \lambda^4}{M^4} \right) \\ - \frac{2M^4}{(4\pi)^2} (N^2-1)^2 \left(1-\ln\frac{M^2}{\bar\mu^2}\right) \;.
	%= -\frac{2(N^2-1)}{Ng^2} \left(1 - \frac38 \frac{Ng^2}{(4\pi)^2} + b_0 \frac{Ng^2}{(4\pi)^2}\right)\lambda^4 + \frac9{26} \frac{N^2-1}{Ng^2} m^4 - \frac{24}{35} \frac{(N^2-1)^2}{Ng^2} M^4 \\
	%-\frac3{4(4\pi)^2} (N^2-1) \left( \frac56 (m^4 - 2\lambda^4) - \frac{m^4+M^4 - 2\lambda^4}2\ln\frac{m^2M^2 + \lambda}{\bar\mu^4} \right. \\ \left. + (m^2+M^2) \sqrt{4\lambda^4-(m^2-M^2)^2}\arctan\frac{\sqrt{4\lambda^4-(m^2-M^2)^2}}{m^2+M^2} + M^4\ln\frac{M^2}{\bar\mu^2} \right) \\ - \frac{2M^4}{(4\pi)^2} (N^2-1)^2 \left(1-\ln\frac{M^2}{\bar\mu^2}\right) \;.
\end{multline}

In the symmetric approach we get instead
\begin{multline}
	\Gamma_\text{sym}(m^2,M^2,\lambda^4) = -\frac{2(N^2-1)}{Ng^2} \left(1 - \frac38 \frac{Ng^2}{(4\pi)^2} + b_0 \frac{Ng^2}{(4\pi)^2}\right)\lambda^4 + \frac9{26} \frac{N^2-1}{Ng^2} m^4 + \frac\beta2 M^4 \\
	-\frac3{4(4\pi)^2} (N^2-1) \left( (m^4 - 2\lambda^4)\left(\frac56 - \frac12\ln\frac{m^2M^2 + \lambda^4}{\bar\mu^4}\right) \right. \\ \left. + (m^2+M^2) \sqrt{4\lambda^4-(m^2-M^2)^2}\arctan\frac{\sqrt{4\lambda^4-(m^2-M^2)^2}}{m^2+M^2} - \frac12 M^4\ln\frac{m^2M^2 + \lambda^4}{M^4} \right) \;.
\end{multline}

In order to determine the free parameters ($b_0$, $\bar\mu^2$, $g^2$, $\beta$) and the zero-temperature condensates ($m_0^2$, $M_0^2$, $\lambda_0^4$), we have the following constraints:
\begin{itemize}
	\item Gribov gap equation $\frac{\partial\Gamma}{\partial\lambda^4} (m_0^2,M_0^2,\lambda_0^4) = 0$,
	\item LCO gap equation for $A^2$ condensate $\frac{\partial\Gamma}{\partial m^2} (m_0^2,M_0^2,\lambda_0^4) = 0$,
	\item LCO gap equation for Gribov ghost condensate $\frac{\partial\Gamma}{\partial M^2} (m_0^2,M_0^2,\lambda_0^4) = 0$,
	\item renormalization group $\frac{(4\pi)^2}{Ng^2} = \frac{11}3 \ln\frac{\bar\mu^2}{\Lambda_{\overline{\text{MS}}}^2}$, with $\Lambda_{\overline{\text{MS}}} = \unit{0.224}{\giga\electronvolt}$ in SU(3) and $\unit{0.331}{\giga\electronvolt}$ in SU(2) \cite{Boucaud:2008gn,Dudal:2017kxb},
	\item two pole masses: $x_0 = \frac12 (m^2+M^2)$, $y_0 = \frac12 \sqrt{4\lambda^4-(m^2-M^2)^2}$.
\end{itemize}
In the $\bar\varphi\varphi$ approach this gives six constraints for six degrees of freedom. In the symmetric approach there is one more free parameter ($\beta$), leaving us with the freedom to choose $\bar\mu^2$ to one of the scales in the logarithms. These scales are not too different from one another; we choose $\bar\mu^2 = \sqrt{m^2M^2 + \lambda^4} = \sqrt{x_0^2+y_0^2}$.

The gluon propagator has poles at the values $p^2_\pm = x_0 \pm iy_0$; in SU(3) we have \cite{Dudal:2018cli} $x_0 = \unit{0.261}{\giga\electronvolt^2}$ and $y_0 = \unit{0.465}{\giga\electronvolt^2}$, and in SU(2) we have \cite{Cucchieri:2011ig} $x_0 = \unit{0.29}{\giga\electronvolt^2}$ and $y_0 = \unit{0.66}{\giga\electronvolt^2}$.

In the $\bar\varphi\varphi$ approach we find \cite{Dudal:2019ing} for SU(3): $b_0 = -3.42$, $\bar\mu = \unit{0.31}{\giga\electronvolt}$; and for SU(2): $b_0 = -1.6$, $\bar\mu = \unit{0.37}{\giga\electronvolt}$.

The symmetric approach is worked out in the \appendixname\ \ref{paramsymm}.

\section{Finite temperature}
To reduce clutter in the subsequent subsections, let us introduce the following shorthands:
\begin{subequations} \label{inot} \begin{gather}
	P^2_\kappa = (2\pi n+\kappa r)^2 T^2 + \vec p^2 \;, \\
	I(\Delta,r,T) = T\int\frac{d^3p}{(2\pi)^3} \ln\left(1-2e^{-\sqrt{\vec p^2+\Delta}/T}\cos r+e^{-2\sqrt{\vec p^2+\Delta}/T}\right) \;, \\
	I(\Delta,0,T) = 2T\int\frac{d^3p}{(2\pi)^3} \ln\left(1-e^{-\sqrt{\vec p^2+\Delta}/T}\right) \;.
\end{gather} \end{subequations}

\subsection{Trace-logarithms}
With a constant background $(\bar A_\mu^a)^h = \delta^{a3} \delta_{\mu0} rT/g$ ($-2\pi<r<2\pi$) in SU(2), we have that
\begin{equation}
	\bar D_\mu^\kappa = \partial_\mu + i\kappa rT\delta_{\mu0} \;,
\end{equation}
where we used the conventions in \appendixname\ \ref{sutwoapp}. As such the eigenvalues of $- \bar D_h^2$ are $P^2_\kappa$. In SU(2), the last two $\tr\ln$'s in \eqref{trlns} thus give the finite-temperature correction
\begin{multline}
	\left(-\frac12-12(1-s)\right) (I(0,r,T) + I(0,0,T) + I(0,-r,T)) + 12(1-s) (I(M^2,r,T) + I(M^2,0,T) + I(M^2,-r,T)) \\
	= \left(-\frac12-12(1-s)\right) \left(2I(0,r,T) - \frac{\pi^2T^2}{45} \right) + 12(1-s) (2I(M^2,r,T) + I(M^2,0,T)) \;,
\end{multline}
where we used the symmetry of $I(\Delta,r,T)$ under $r\to-r$.

In SU(3), charge conjugation invariance implies \cite{Kroff:2018ncl} it is enough to consider the background $(\bar A_\mu^a)^h = \delta^{a3} \delta_{\mu0} rT/g$ ($-2\pi<r<2\pi$). With the conventions in \appendixname\ \ref{suthreeapp}, $\bar D_\mu^h$ evaluates to:
\begin{subequations} \begin{align}
	\mx v_{3,8} \;: & \quad \partial_\mu \;, \\
	\mx v_1^\pm \;: & \quad \partial_\mu \pm irT\delta_{\mu0} \;, \\
	\mx v_2^\pm \;: & \quad \partial_\mu \pm \tfrac i2rT\delta_{\mu0} \;, \\
	\mx v_3^\pm \;: & \quad \partial_\mu \mp \tfrac i2rT\delta_{\mu0} \;.
\end{align} \end{subequations}
This allows us to compute the finite-temperature correction to the last two $\tr\ln$'s in \eqref{trlns} in SU(3):
\begin{multline}
	\left(-\frac12-32(1-s)\right) \left(2I(0,0,T) + I(0,r,T) + I(0,-r,T) + 2I(0,\tfrac r2,T) + 2I(0,-\tfrac r2,T)\right) \\ + 32(1-s) \left(2I(M^2,0,T) + I(M^2,r,T) + I(M^2,-r,T) + 2I(M^2,\tfrac r2,T) + 2I(M^2,-\tfrac r2,T)\right) \\
	= 2\left(-\frac12-32(1-s)\right) \left(- \frac{\pi^2T^2}{45} + I(0,r,T) + 2I(0,\tfrac r2,T)\right) + 64(1-s) \left(I(M^2,0,T) + I(M^2,r,T) + 2I(M^2,\tfrac r2,T)\right) \;.
\end{multline}

The gluon trace-logarithm (the first trace in the last line of \eqref{trlns}) is more complicated. In the $\bar A^h$ approach, the Gribov term is, at finite temperature, replaced with
\begin{equation}
	\delta_{\mu\nu} \delta^{ab} \frac{\lambda^4}{p^2+M^2} \to \delta_{\mu\nu} \frac{\lambda^4}N f^{ace} \left(\frac1{P^2+M^2}\right)^{cd} f^{dbe} \;.
\end{equation}
To evaluate this, we use \eqref{epsilons} for SU(2) and \eqref{operatorfof} for SU(3).

In SU(2), the eigenvalues of the quadratic gluon operator (the analogon of the first term of the last line of \eqref{trlns}) are
\begin{equation} \begin{gathered}
	P^2_\pm + m^2 + \frac{\lambda^4}2 \left( \frac1{P^2_\pm + M^2} + \frac1{P^2_0 + M^2} \right) \;\text{, and} \\
	P^2_0 + m^2 + \frac{\lambda^4}2 \left( \frac1{P^2_{+1} + M^2} + \frac1{P^2_{-1} + M^2} \right) \;.
\end{gathered} \end{equation}
For the trace-logarithm, this gives $\frac{d-1}2$ times
\begin{multline}
	\ln \left((P^2_\pm + m^2)(P^2_\pm + M^2)(P^2_0 + M^2) + \frac{\lambda^4}2 (P^2_\pm + M^2 + P^2_0 + M^2)\right) \\ + \ln\left((P^2_0 + m^2)(P^2_{+1} + M^2)(P^2_{-1} + M^2) + \frac{\lambda^4}2 (P^2_{+1} + M^2 + P^2_{-1} + M^2)\right) \\ - 2\ln(P^2_0 + M^2) - 2\ln(P^2_\pm + M^2) \;,
\end{multline}
where the indices ``$\pm$'' need to be summed over. The terms on the last line give (after multiplication with $\frac{d-1}2$ and taking the trace)
\begin{equation}
	-9 \tr_{T=0}\ln(-\partial^2 + M^2) - 6I(M^2,r,T) - 3I(M^2,0,T) \;.
\end{equation}
What is left are three sixth-order polynomials in $n$.\footnote{The second one, from the $r=0$ state, is actually a third-order polynomial in $n^2$, which can be factored, but handling this one numerically as well saves handwork and does not waste relatively that much more time doing numerics.} In order to deal with them, we use \eqref{finitetmaster}. This is straightforward to implement numerically, but does considerably slow down the computations.

In SU(3), the eigenvalues of the gluon propagator are
\begin{subequations} \begin{align}
	\mx v_3 \;: & \quad P_0^2+m^2 + \frac{\lambda^4}3 \left( \frac1{P_{+1}^2+M^2} + \frac1{P_{-1}^2+M^2} + \tfrac12 \frac1{P_{+1/2}^2+M^2} + \tfrac12 \frac1{P_{-1/2}^2+M^2} \right) \;, \\
	\mx v_8 \;: & \quad P_0^2+m^2 + \frac{\lambda^4}2 \left( \frac1{P_{+1/2}^2+M^2} + \frac1{P_{-1/2}^2+M^2} \right) \;, \\
	\mx v_1^\pm \;: & \quad P_{\pm1}^2+m^2 + \frac{\lambda^4}3 \left( \frac1{P_0^2+M^2} + \frac1{P_{\pm1}^2+M^2} + \frac1{P_{\pm1/2}^2+M^2} \right) \;, \\
	\mx v_2^\pm, \mx v_3^\mp \;: & \quad P_{\pm1/2}^2+m^2 + \frac{\lambda^4}3 \left( \frac1{P_0^2+M^2} + \tfrac12\frac1{P_{\pm1}^2+M^2} + \frac1{P_{\pm1/2}^2+M^2} + \tfrac12\frac1{P_{\mp1/2}^2+M^2} \right) \;.
\end{align} \end{subequations}
The trace-logarithm now gives polynomials up to tenth order, for which we again use \eqref{finitetmaster}, and the denominators lead to the subtraction
\begin{multline}
	-\frac{d-1}2 \Bigg( 6\tr\ln(P_0^2+M^2) + 4\tr\ln(P_{\pm1}^2+M^2) + 7\tr\ln(P_{\pm\frac12}^2+M^2) \Bigg) \\ = -42\tr_{T=0}\ln(-\partial^2 + M^2) - 9I(M^2,0,T) - 12I(M^2,r,T) - 21I(M^2,\tfrac r2,T) \;.
\end{multline}

In the Kroff--Reinosa approach, the Gribov term is, at finite temperature, replaced with
\begin{equation}
	\delta_{\mu\nu} \delta^{ab} \frac{\lambda^4}{p^2+M^2} \to \delta_{\mu\nu} \delta^{ab} \frac{\lambda^4}{P^2+M^2} \;.
\end{equation}
This gives instead: For SU(2) $\frac{d-1}2$ times
\begin{equation}
	\ln \left(P^2_\pm + m^2 + \frac{\lambda^4}{P^2_\pm + M^2}\right) + \ln\left(P^2_0 + m^2 + \frac{\lambda^4}{P^2_0 + M^2}\right) \;,
\end{equation}
and for SU(3) $\frac{d-1}2$ times
\begin{equation}
	2\ln\left(P^2_0 + m^2 + \frac{\lambda^4}{P^2_0 + M^2}\right) + \ln \left(P^2_\pm + m^2 + \frac{\lambda^4}{P^2_\pm + M^2}\right) + 2\ln \left(P^2_{\pm\frac12} + m^2 + \frac{\lambda^4}{P^2_{\pm\frac12} + M^2}\right) \;.
\end{equation}
To compute this, we see that
\begin{subequations} \begin{multline}
	\tr\ln \left(P^2_r + m^2 + \frac{\lambda^4}{P^2_r + M^2}\right) = \tr\ln(P^2_r+z_+) + \tr\ln(P^2_r+z_-) - \tr\ln(P^2_r+M^2) \\ = \tr_{T=0} \ln \left(p^2 + m^2 + \frac{\lambda^4}{p^2 + M^2}\right) + I(z_+,r,T) + I(z_-,r,T) - I(M^2,r,T) \;,
\end{multline}
where
\begin{equation}
	z_\pm = \frac12 \left(m^2+M^2 \pm i\sqrt{4\lambda^4-(m^2-M^2)^2}\right) \;.
\end{equation} \end{subequations}

\subsection{Extremization}
Once we have computed the effective action, we solve the gap equation to find the Gribov parameter $\lambda$ and minimize with respect to the condensates. The Gribov gap equation corresponds to finding a \emph{maximum}, which means the final solution will be a saddle point in the four-dimensional space of the parameters. This complicates numerical minimization.

In order to find this saddle point, we found it most straightforward to use iteration. Starting from seed values for the parameters (obtained from extrapolating from previous data obtained at, for example, lower temperature), we first maximize with respect to the Gribov parameter, then minimize with respect to the other parameters, maximize with respect to the Gribov parameter again, etc. until successive steps do not lead to significant changes any longer. Then we move on to the next value of the temperature.

This iteration is sometimes unstable, and may diverge. We found this can be cured by ``damping'' the change in the Gribov parameter $\lambda$ in successive steps. If $\lambda_\text{o}^2$ is the previous value (of the square) and $\lambda_\text{n}^2$ the newly obtained one, we use
\begin{equation}
	\frac{a\lambda_\text{o}^2+\lambda_\text{n}^2}{a+1}
\end{equation}
for the next value of $\lambda^2$. Taking $a=1$ often leads to fast convergence for low temperatures. In the deconfined phases, taking $a=10$ or some such generally ensures convergence.

\subsection{Results in $\bar\varphi\varphi$ case}
\begin{figure}[t]
	\begin{subfigure}{0.35\textwidth} \includegraphics[width=\textwidth]{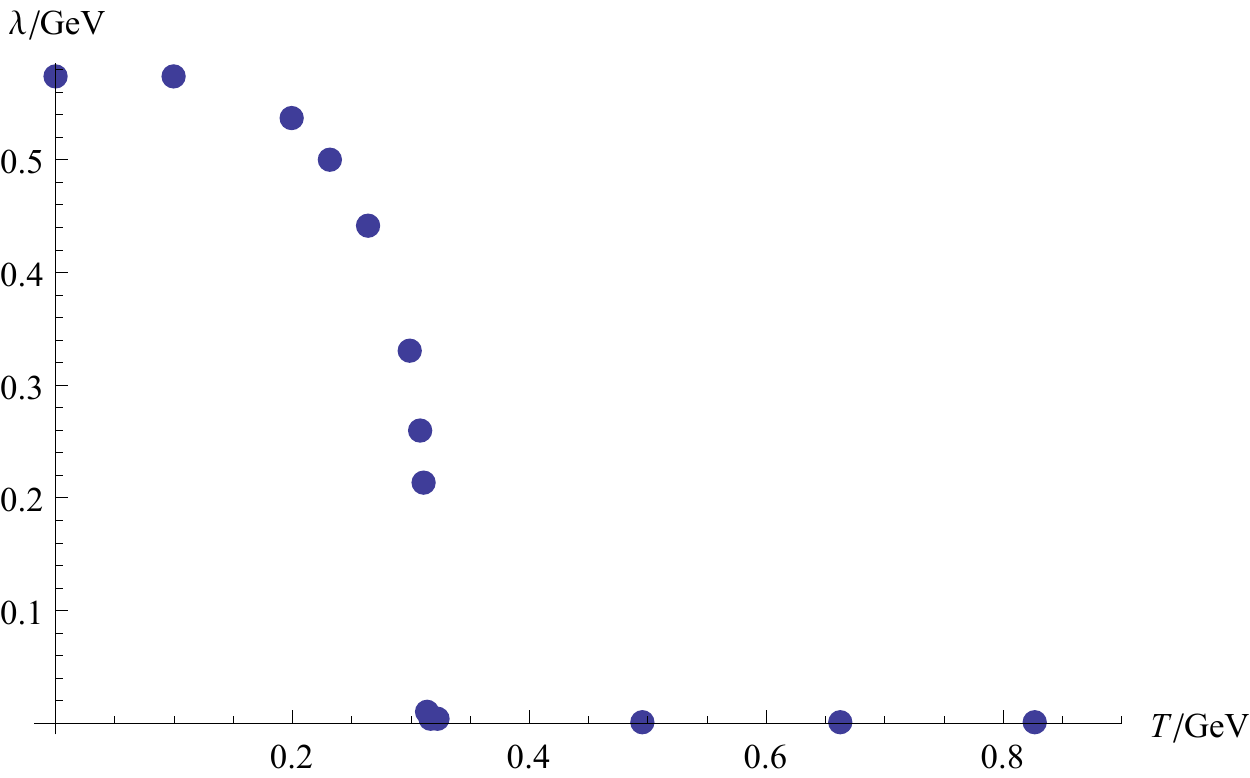} \caption{\label{geensusyoponzemanierlambda}} \end{subfigure} \hfill \begin{subfigure}{0.35\textwidth} \includegraphics[width=\textwidth]{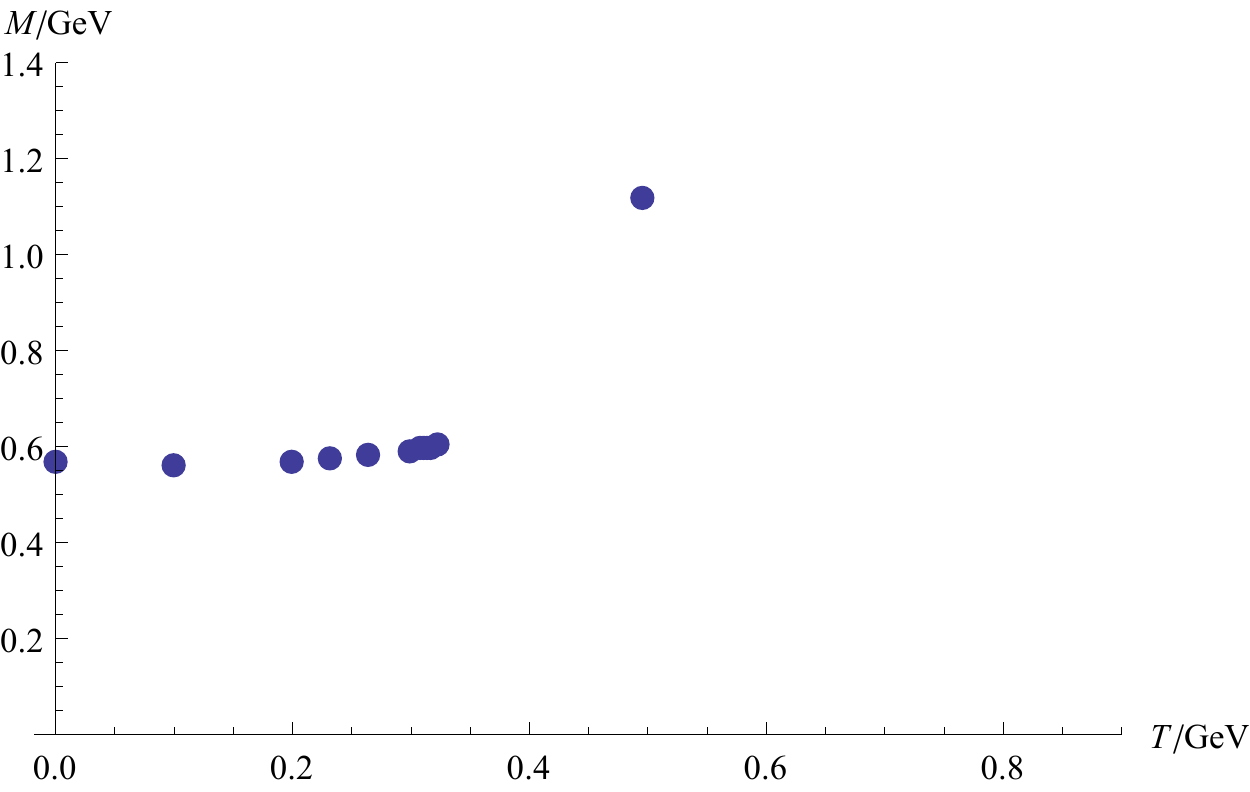} \caption{\label{geensusyoponzemanierM}} \end{subfigure} \hfill \begin{subfigure}{0.25\textwidth} \includegraphics[width=\textwidth]{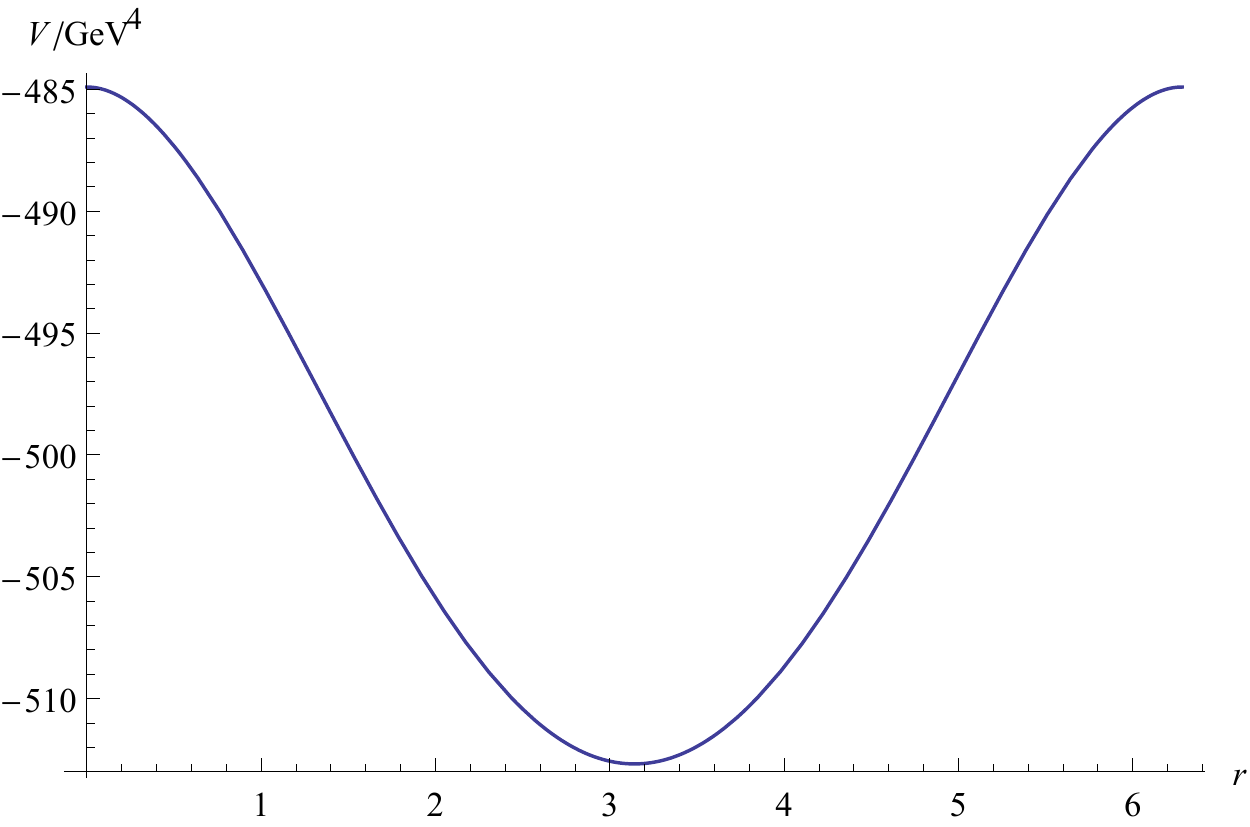} \caption{\label{geensusyoponzemanierV}} \end{subfigure}
	\caption{Some of the results obtained in the $\bar A^h$ approach for the $\bar\varphi\varphi$ case in SU(2). As the numerics are quite heavy, we did the computation for a smaller selection of temperatures. \figurename\ \subref{geensusyoponzemanierlambda} shows the Gribov parameter $\lambda$, which goes to zero at $T \approx \unit{0.32}{\giga\electronvolt}$. Not shown is the Polyakov loop $r$, which is equal to $\pi$ throughout. \figurename\ \subref{geensusyoponzemanierM} shows how the $\langle\bar\varphi\varphi\rangle$ condensate (proportional to the mass parameter $M^2$) starts a rapid increase after $\lambda$ has gone to zero. (Points for temperatures beyond \unit{0.50}{\giga\electronvolt} fall outside the plot.) Also not shown is $m$, which does not vary all that much in the temperature range shown. \figurename\ \subref{geensusyoponzemanierV} finally shows the potential of the Polyakov loop $r$ (keeping the other parameters fixed to the values they have in the minimum of the potential) for $T=\unit{1.3}{\giga\electronvolt}$, showing clearly that $r=\pi$ is still the minimum. \label{geensusyoponzemanier}}
\end{figure}

\begin{figure}[t]
	\begin{subfigure}{0.35\textwidth} \includegraphics[width=\textwidth]{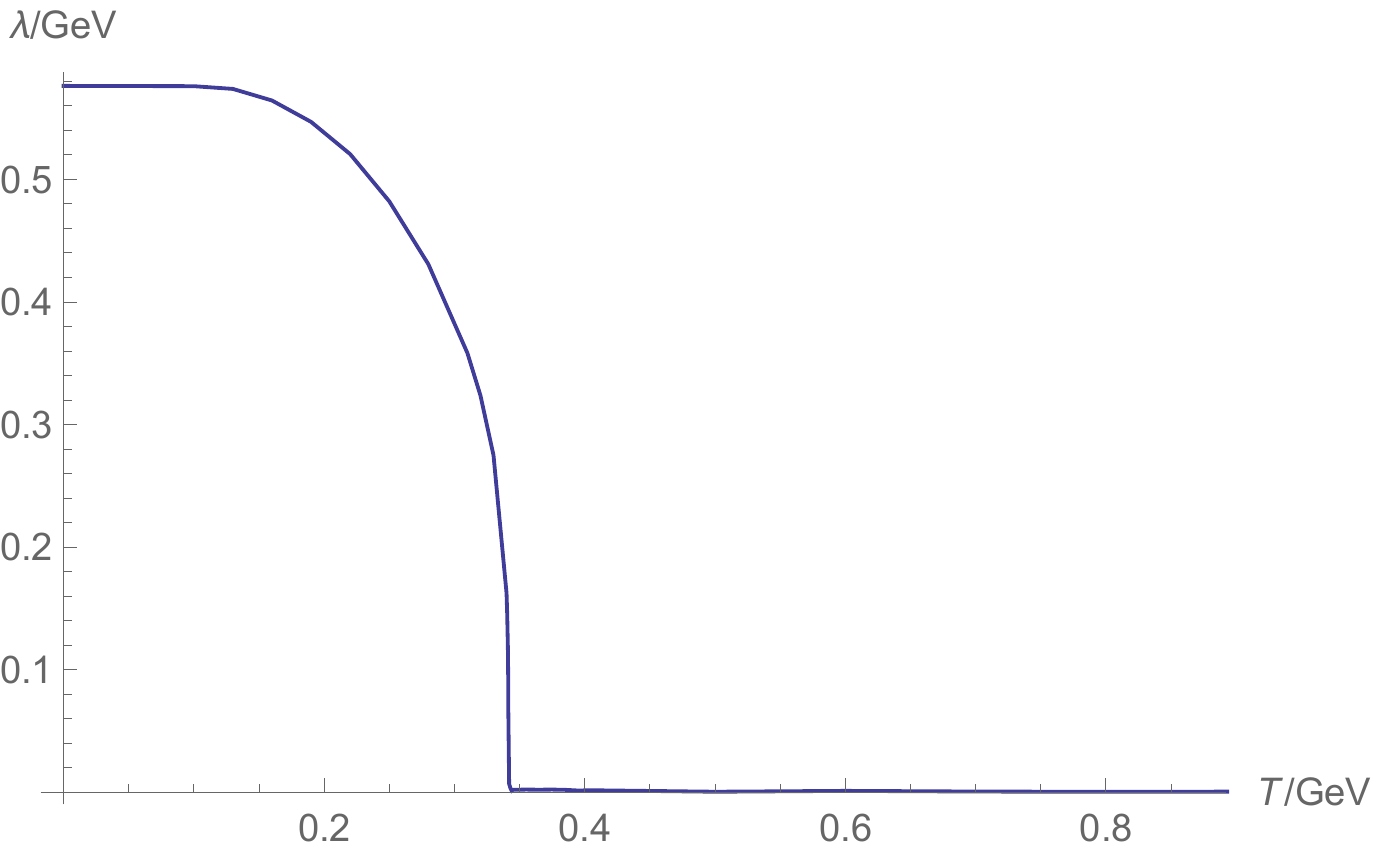} \caption{\label{geensusyopKRmanierlambda}} \end{subfigure} \hfill \begin{subfigure}{0.35\textwidth} \includegraphics[width=\textwidth]{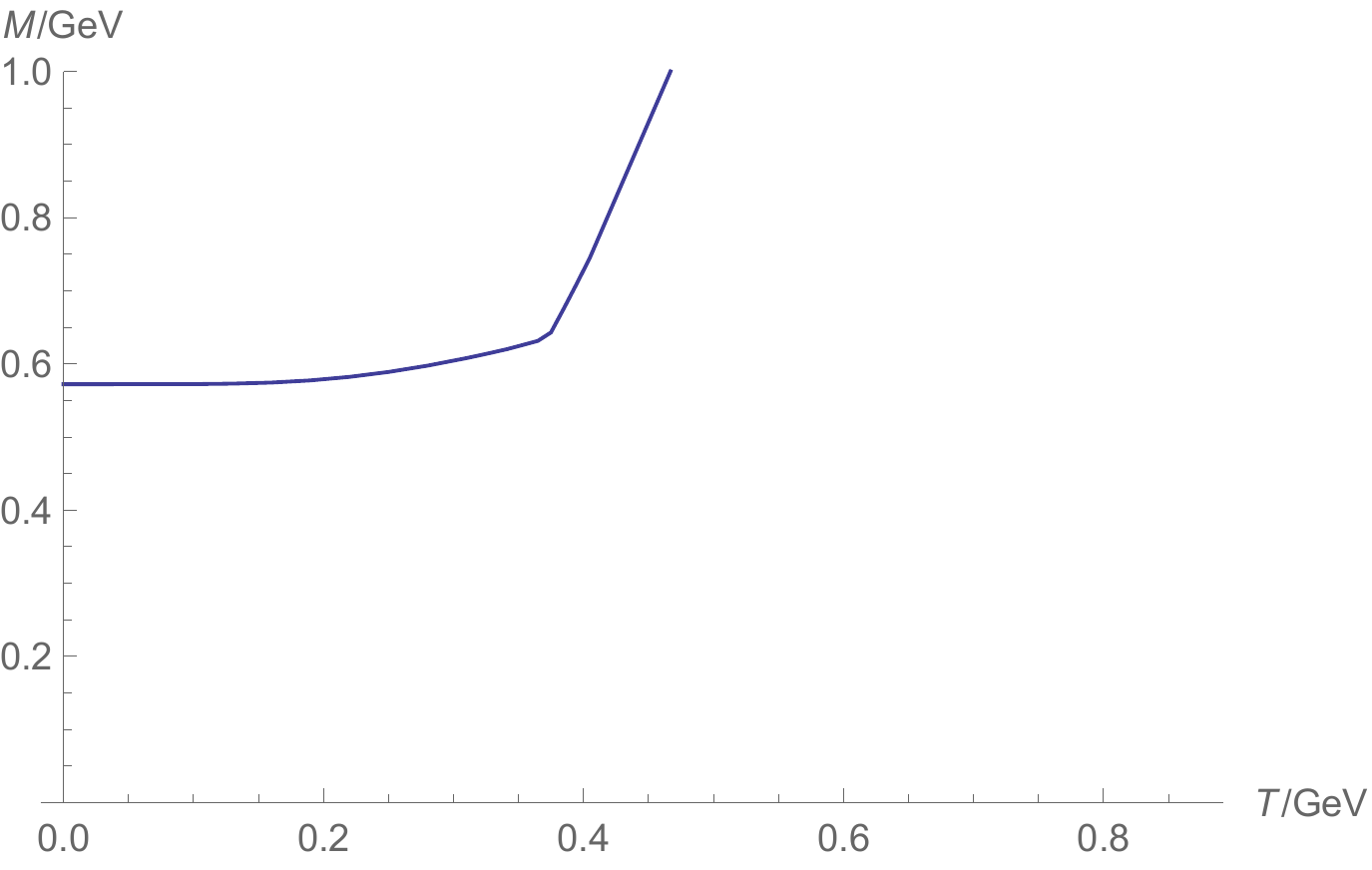} \caption{\label{geensusyopKRmanierM}} \end{subfigure} \hfill \begin{subfigure}{0.25\textwidth} \includegraphics[width=\textwidth]{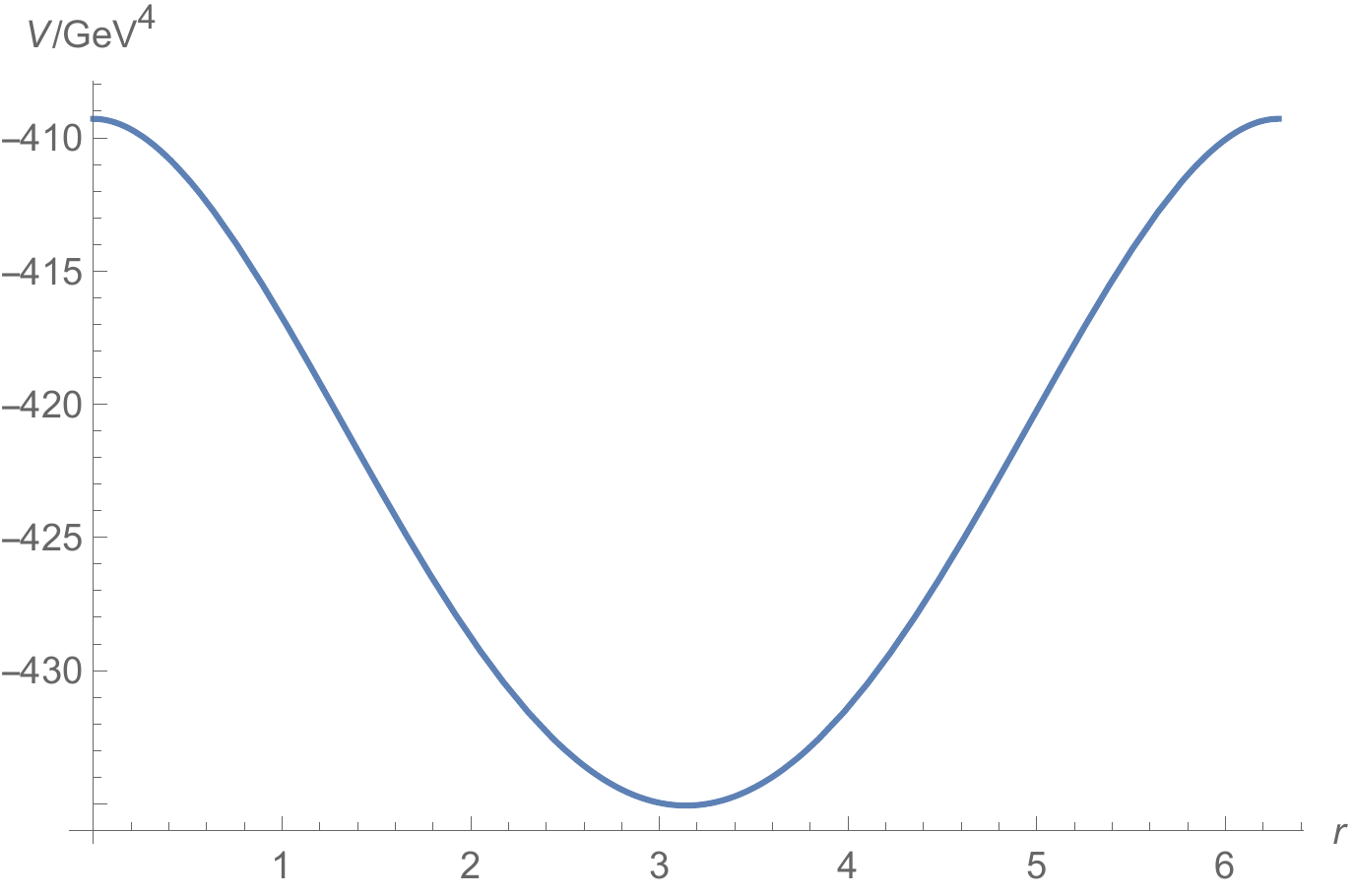} \caption{\label{geensusyopKRmanierV}} \end{subfigure}
	\caption{Some of the results obtained in the KR approach for the $\bar\varphi\varphi$ case in SU(2). \figurename\ \subref{geensusyopKRmanierlambda} shows the Gribov parameter $\lambda$, which goes to zero at $T \approx \unit{0.34}{\giga\electronvolt}$. Not shown is the Polyakov loop $r$, which is equal to $\pi$ throughout. \figurename\ \subref{geensusyopKRmanierM} shows how the $\langle\bar\varphi\varphi\rangle$ condensate (proportional to the mass parameter $M^2$) starts a rapid increase after $\lambda$ has gone to zero. Also not shown is $m$, which does not vary all that much in the temperature range shown. \figurename\ \subref{geensusyopKRmanierV} finally shows the potential of the Polyakov loop $r$ (keeping the other parameters fixed to the values they have in the minimum of the potential) for $T=\unit{1.3}{\giga\electronvolt}$, showing clearly that $r=\pi$ is still the minimum. \label{geensusyopKRmanier}}
\end{figure}

With the $\bar\varphi\varphi$ approach in SU(2) with the Polyakov loop in the $\bar A^h$ approach we still did not find any phase transition even at $T=\unit{1.3}{\giga\electronvolt}$ (see \figurename\ \ref{geensusyoponzemanierV}),\footnote{In the $\bar\varphi\varphi$ approach, the renormalization scale $\bar\mu$ is usually held fixed to its zero temperature value. For temperatures higher than this value, we took $\bar\mu = T$ instead, but keeping $\bar\mu$ fixed did not give qualitatively different results.} while $\lambda$ goes to zero around \unit{0.32}{\giga\electronvolt} (see \figurename\ \ref{geensusyoponzemanierlambda}). In the KR approach the same happens: $\lambda$ goes to zero around \unit{0.34}{\giga\electronvolt} (see \figurename\ \ref{geensusyopKRmanierlambda}), while the Polyakov loop still signals confinement around $T = \unit{1.3}{\giga\electronvolt}$ (see \figurename\ \ref{geensusyopKRmanierV}).

This shows that the Gribov parameter is not really an order parameter for confinement in this case. The discrepancy is due to the difference in mass between $\varphi$ and $\omega$: these fields are supposed to have their determinants cancel, which does not happen here. If these determinants were to cancel, $\lambda\to0$ would bring us back to the Curci--Ferrari-type model considered in \cite{Dudal:2022nnu}, where confinement is recovered for $T=\unit{0.32}{\giga\electronvolt}$. Without this cancellation of the two determinants, $M^2$ increases without bound (see \figurename\ \ref{geensusyoponzemanierM} and \ref{geensusyopKRmanierM}) (while $m^2$ shows only a modest increase) and this seems to drive $r$ to $\pi$.

To conclude, it appears that the $\bar\varphi\varphi$ case is flawed and does not describe the physics well. Due to these shortcomings, we did not bother to investigate the (more involved) SU(3) theory.

\subsection{Results in symmetric case: $\bar A^h$ approach}
\begin{figure}[t]
	\begin{subfigure}{0.35\textwidth} \includegraphics[width=\textwidth]{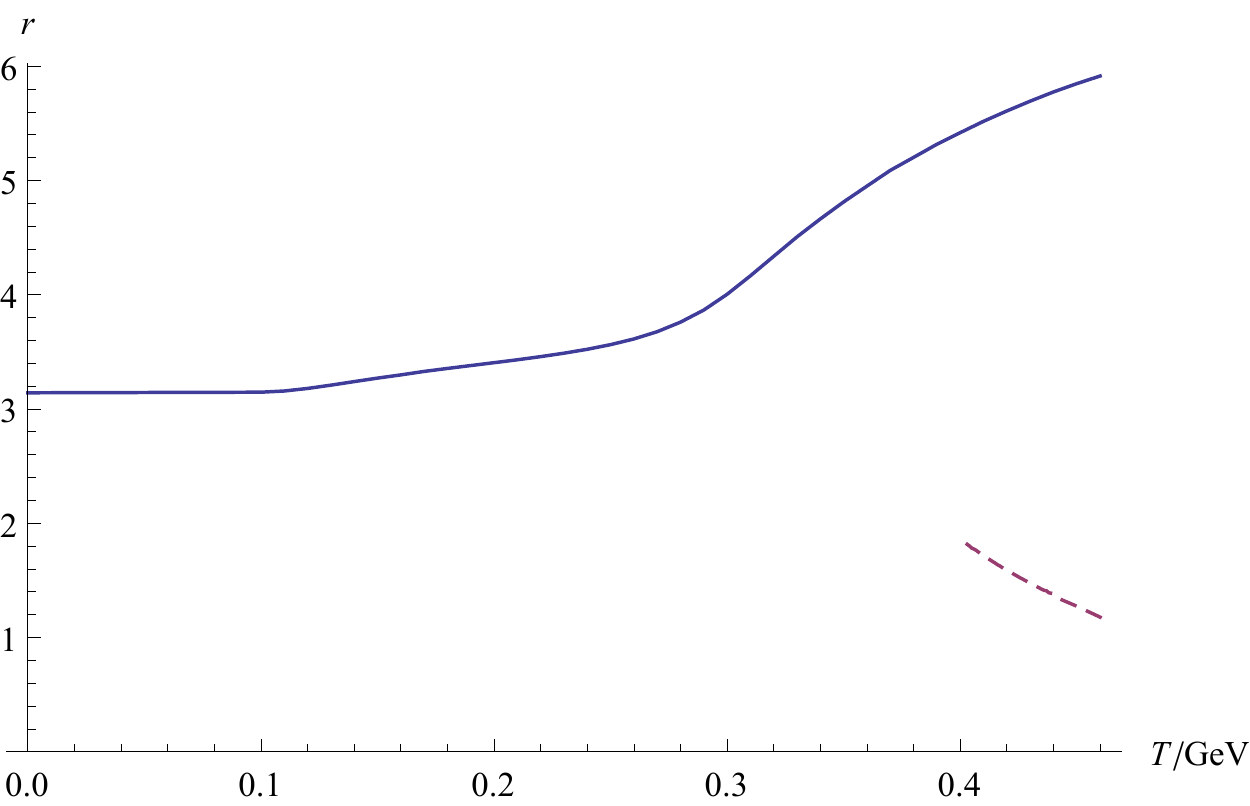} \caption{\label{susyoponzemanierr2}} \end{subfigure} \qquad \begin{subfigure}{0.35\textwidth} \includegraphics[width=\textwidth]{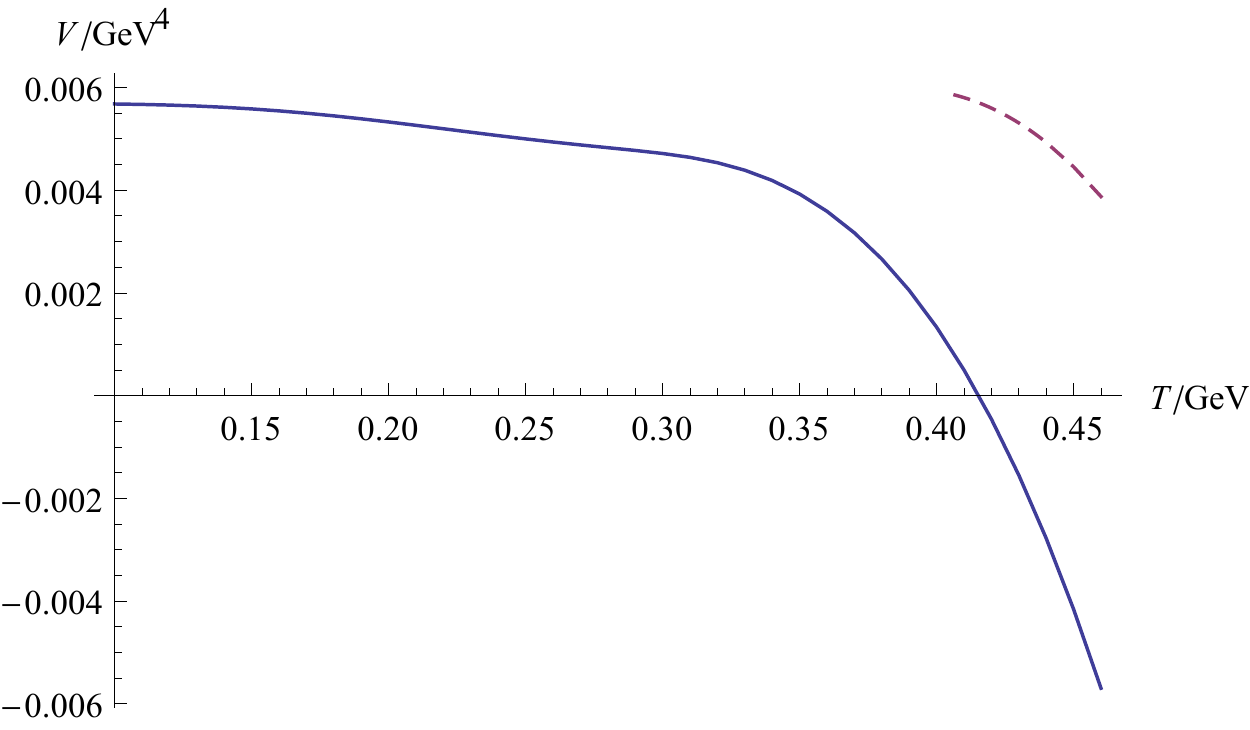} \caption{\label{susyoponzemanierV2}} \end{subfigure}
	\caption{Some of the results obtained in the $\bar A^h$ approach for the symmetric case in SU(2) for the two minima we found: the ``overconfining'' minimum ($r>\pi$) in a full line and the deconfining minimum in a dashed line. Not shown are Gribov parameter and the dimension-two condensates, which do not vary much and also do not differ much between the two vacua. \figurename\ \subref{susyoponzemanierr2} shows the Polyakov loop $r$ as a function of temperature. Already at very small temperature, $r>\pi$, which implies the quark free energy has an imaginary part. \figurename\ \subref{susyoponzemanierV2} shows the energy in the minima. The ``overconfining'' vacuum is preferred for the entire temperature range. \label{susyoponzemanier2}}
\end{figure}

\begin{figure}[t]
	\begin{subfigure}{0.3\textwidth} \includegraphics[width=\textwidth]{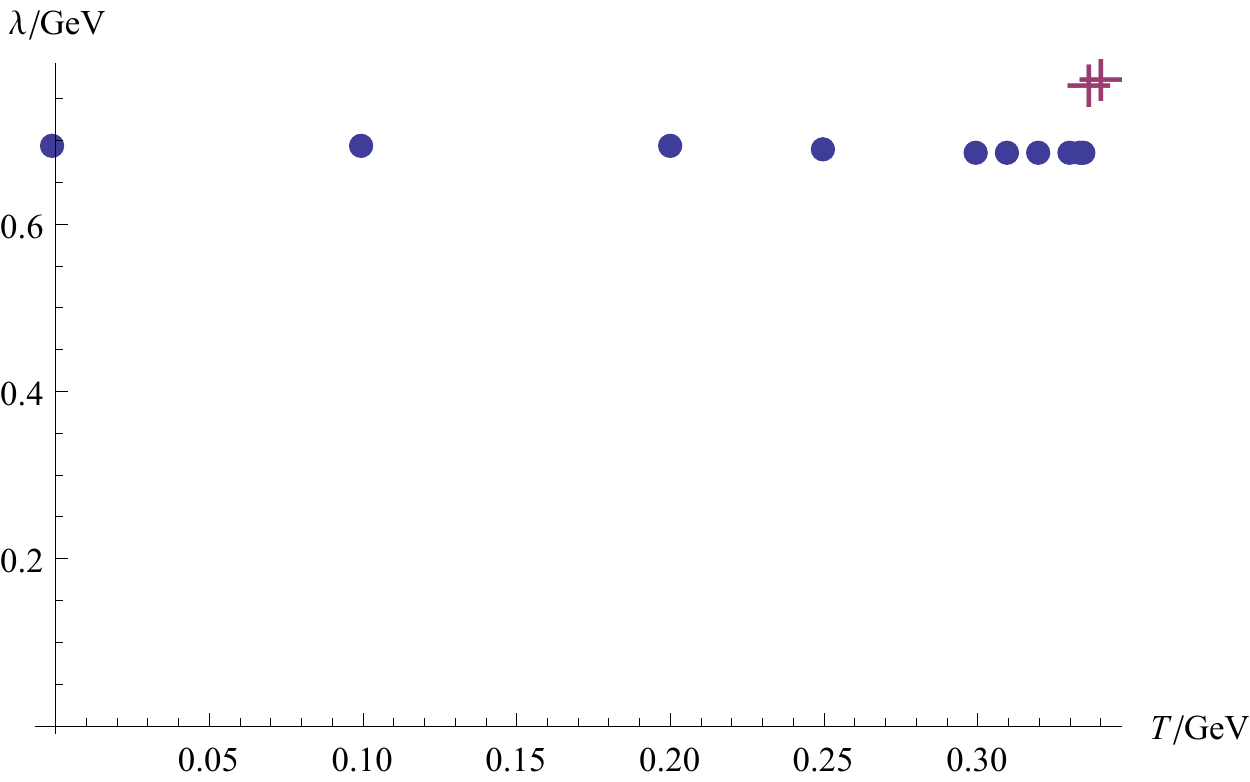} \caption{\label{susyoponzemanierlambda3}} \end{subfigure} \hfill \begin{subfigure}{0.3\textwidth} \includegraphics[width=\textwidth]{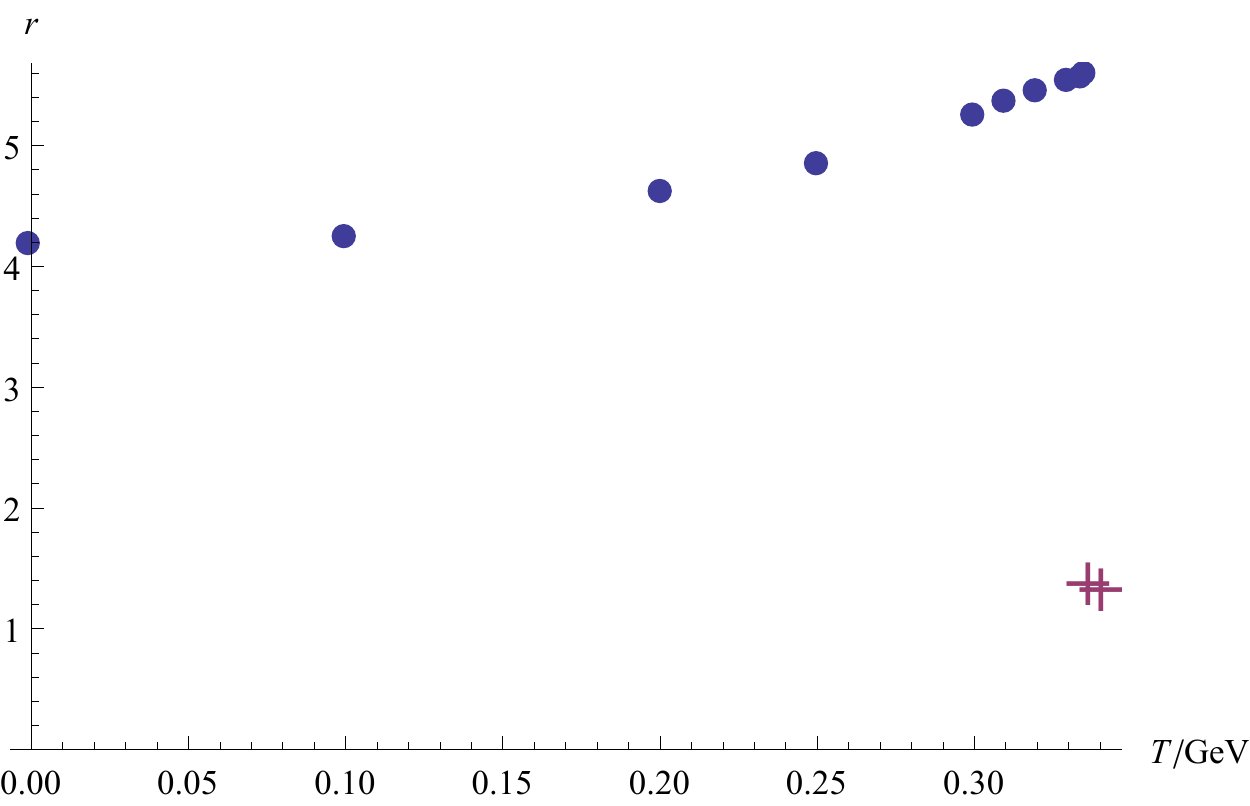} \caption{\label{susyoponzemanierr3}} \end{subfigure} \hfill \begin{subfigure}{0.35\textwidth} \includegraphics[width=\textwidth]{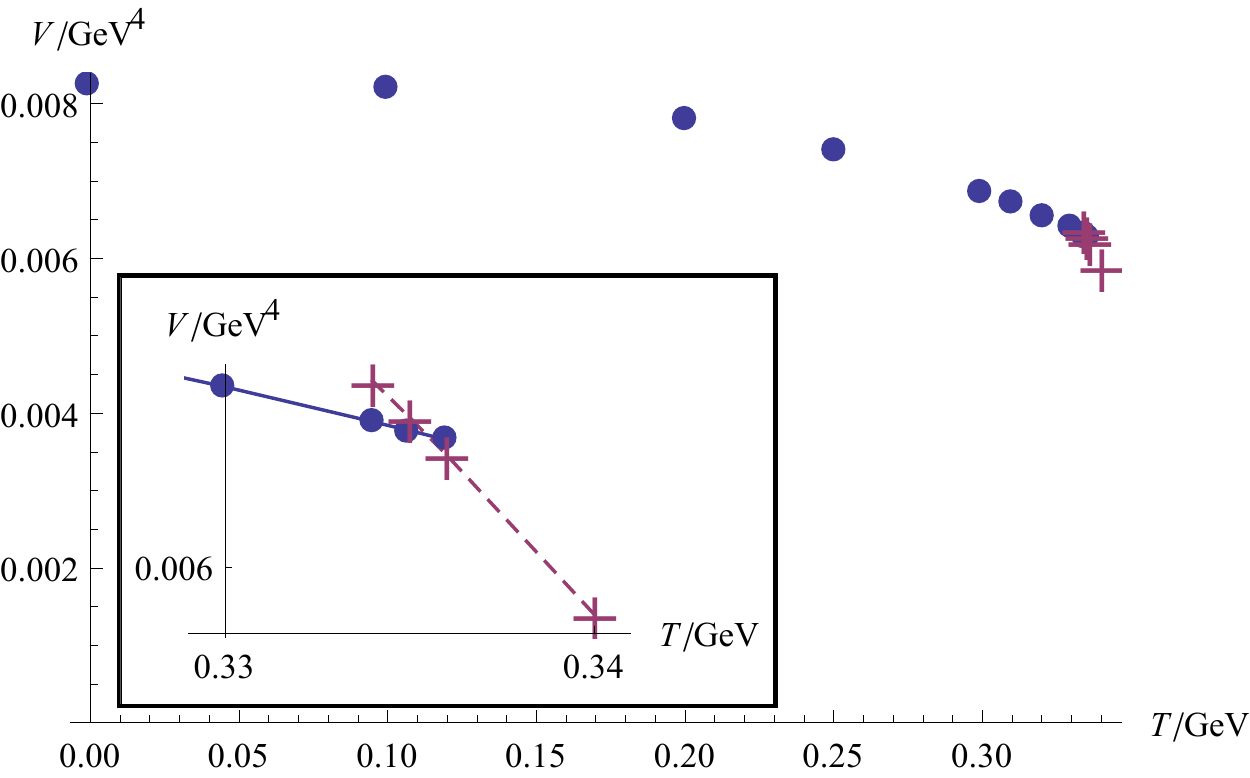} \caption{\label{susyoponzemanierV3}} \end{subfigure}
	\caption{Some of the results obtained in the $\bar A^h$ approach for the symmetric case in SU(3) for the two minima we found: the ``overconfining'' minimum ($r > 4\pi/3 = 4.19$) with dots and the deconfining minimum in plus signs. Again the numerics are quite heavy, so we did the computation for a smaller selection of temperatures. \figurename\ \subref{susyoponzemanierlambda3} shows the Gribov parameter, which jumps to higher values when entering the deconfined phase. Not shown are the dimension-two condensates, which do not vary much and also do not differ much between the two vacua. \figurename\ \subref{susyoponzemanierr3} shows the Polyakov loop $r$ as a function of temperature. Already at very small temperatures, $r > 4\pi/3$, which implies the quark free energy has an imaginary part. \figurename\ \subref{susyoponzemanierV3} shows the energy in the minima with inset zoomed in on the transition. \label{susyoponzemanier3}}
\end{figure}

In the $\bar A^h$ approach for the symmetric case, the determinants of the $\varphi$ and $\omega$ propagators cancel, such that $r$ is not constant anymore. It turns out, however, that $r$ starts increasing in value the moment temperature is switched on, see \figurename\ \ref{susyoponzemanierr2} and \ref{susyoponzemanierr3}. A value of $r$ \emph{higher} than its confining value (called ``overconfining'' in the following) suggests the Polyakov loop itself is negative, or the quark free energy has an imaginary part.

For SU(2), this overconfining minimum persist for all the temperature values we investigated. For $T > \unit{0.40}{\giga\electronvolt}$, we found a second ``normal'' deconfining solution. However, the energy in this minimum remains higher than the energy in the overconfining minimum, and the situation shows no signs of improving with increasing temperature, see \figurename\ \ref{susyoponzemanierV2}. Given the difficulty of finding this deconfining minimum, we cannot rule out the existence of additional minima. The second-order phase transition one expects in SU(2), where the confining minimum spontaneously ``rolls'' into the deconfining minimum, certainly does not happen though.

For SU(3) as well, the Polyakov loop does not remain in its symmetric point $r = 4\pi/3$ already at low temperatures, see \figurename\ \ref{susyoponzemanierr3}. Instead it goes up to $5.58$ at $T = \unit{0.335}{\giga\electronvolt}$. This time we do find a transition at $T_c = \unit{0.335}{\giga\electronvolt}$, see \figurename\ \ref{susyoponzemanierV3}, and $r$ is good and well below $4\pi/3$ after the transition, signaling deconfinement. The Gribov parameter $\lambda$ goes \emph{up} when going through the transition, as seen in \figurename\ \ref{susyoponzemanierlambda3}.

We can conclude that the $\bar A^h$ approach also has some flaws, indicated by the Polyakov loop $r$ increasing in value rather than staying constant during what we would expect to be the confining phase. Furthermore we did not find any deconfined phase for SU(2) in the temperature range we investigated (until $T = \unit{0.46}{\giga\electronvolt}$), and the trends in the vacuum energies do not suggest a deconfined phase will soon be found for higher temperatures. Finally, the transition we did find for SU(3) is at a temperature much higher than found in other works. A lattice computation (see Table 6 in \cite{Lucini:2003zr}, taking for the string tension a typical value of $\sqrt{\sigma}=\unit{0.44}{\giga\electronvolt}$, see \cite{Bali:2000gf} for more details) gives $T_c = \unit{0.28}{\giga\electronvolt}$; other approaches usually find even lower values, see Table 6.1 in \cite{Reinosa:2020mnx} for a selection.

One might speculate that the fact the above results are deviating so from what is expected, is related to the observation made in \cite{Kroff:2018ncl}: in principle, when we go on-shell in the $h$-sector via the $\tau$-equation of motion, the $h$-field must evidently be periodic, but up to a $\mathbb{Z}_N$ twist. As of now, we have not been able to find a way to deal with the twisted sectors in the path integral, and we must restrict ourselves to a fixed twist sector.

\subsection{Results in symmetric case: KR approach}
\begin{figure}[t]
	\begin{subfigure}{0.35\textwidth} \includegraphics[width=\textwidth]{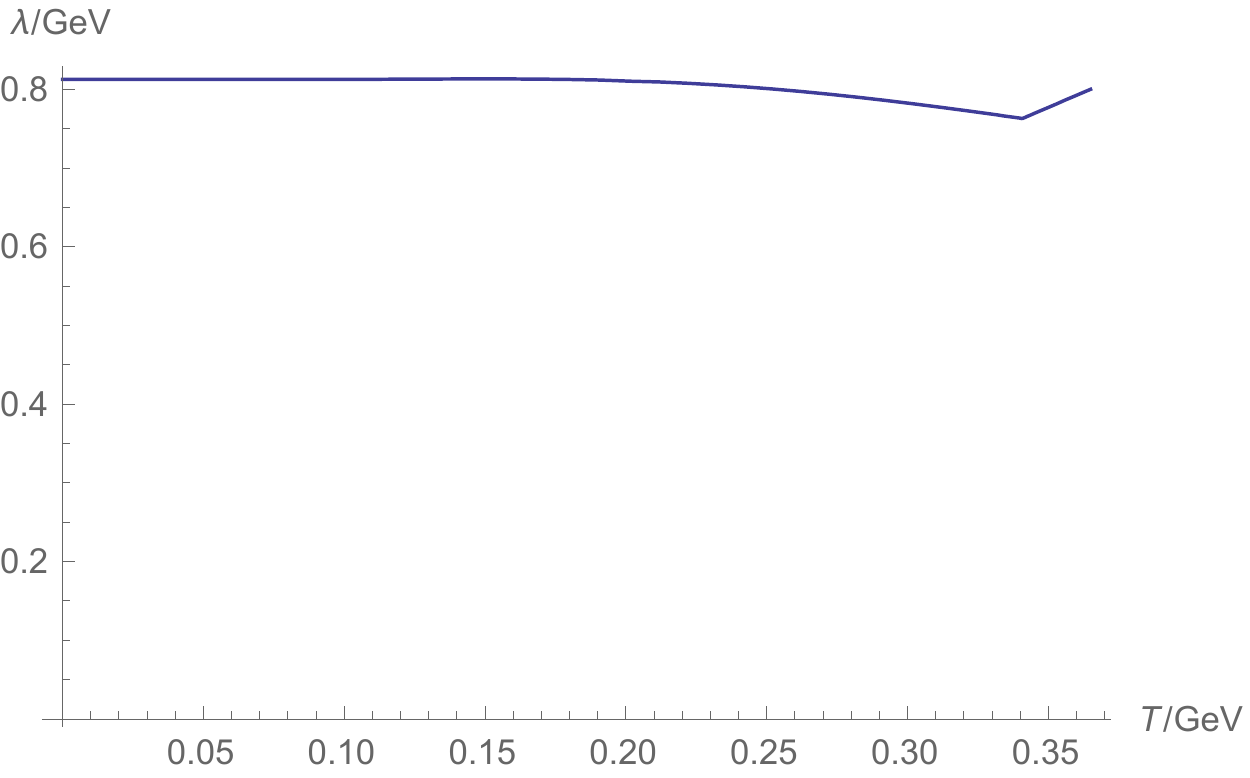} \caption{\label{susyopKRmanierlambda2}} \end{subfigure} \qquad \begin{subfigure}{0.35\textwidth} \includegraphics[width=\textwidth]{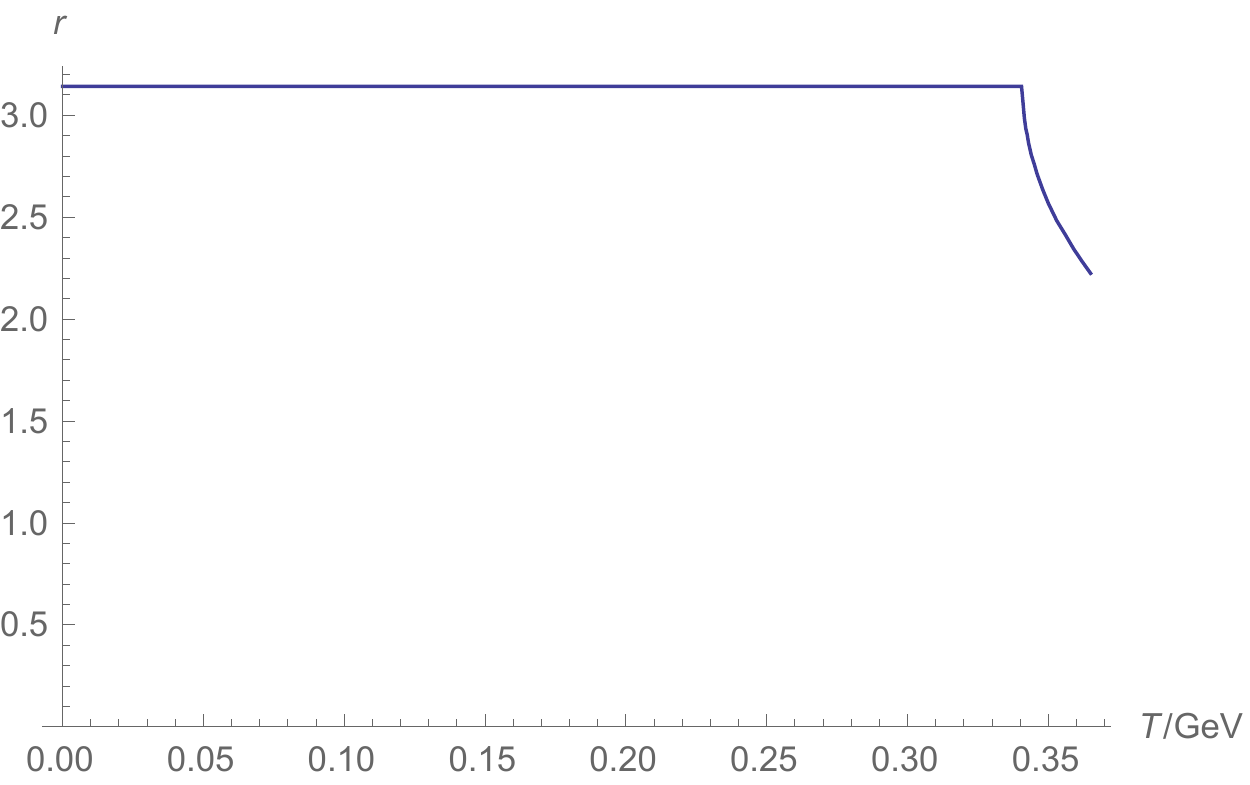} \caption{\label{susyopKRmanierr2}} \end{subfigure}
	\caption{Some of the results obtained in the KR approach for the symmetric case in SU(2). Not shown are the dimension-two condensates, which do not vary much and also do not change much through the transition. \figurename\ \subref{susyopKRmanierlambda2} shows the Gribov parameter $\lambda$ and \figurename\ \subref{susyopKRmanierr2} shows the Polyakov loop $r$ as a function of temperature. The second-order transition at $T_c = \unit{0.34}{\giga\electronvolt}$ is clear in the sudden drop in $r$. \label{susyopKRmanier2}}
\end{figure}

\begin{figure}[t]
	\begin{subfigure}{0.3\textwidth} \includegraphics[width=\textwidth]{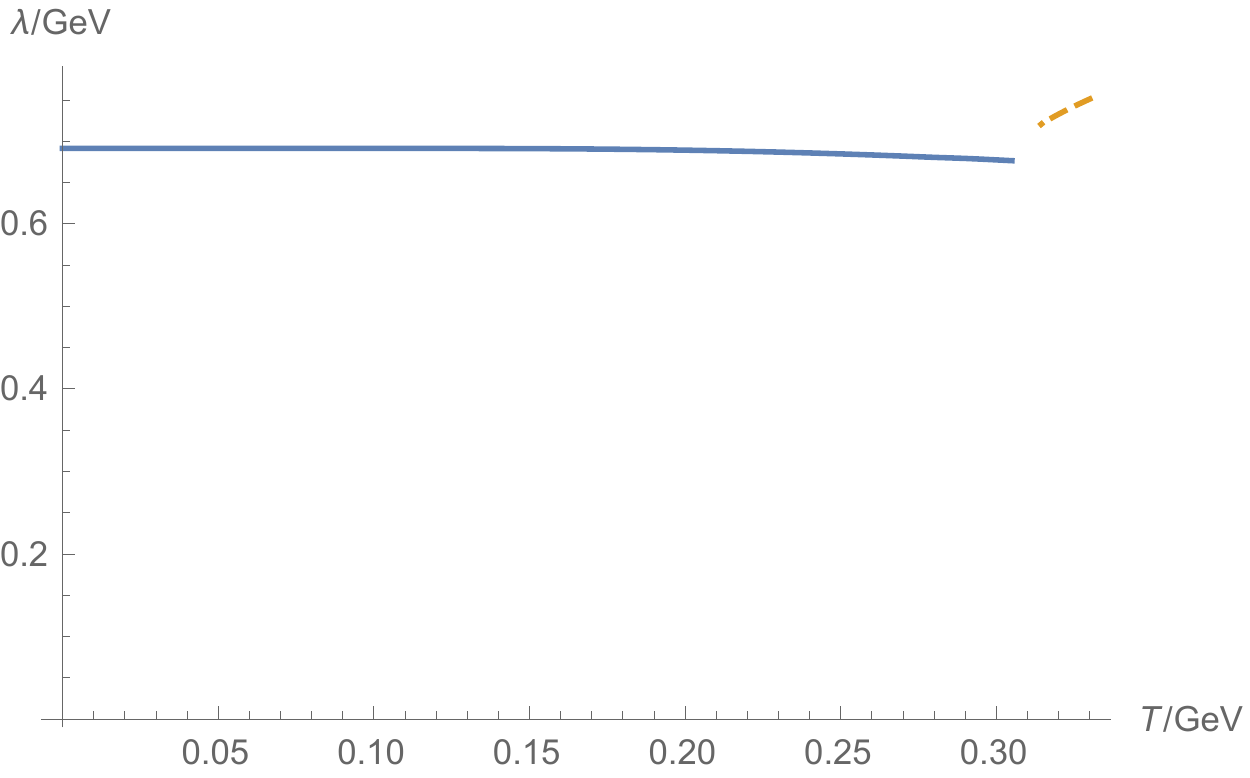} \caption{\label{susyopKRmanierlambda3}} \end{subfigure} \hfill \begin{subfigure}{0.3\textwidth} \includegraphics[width=\textwidth]{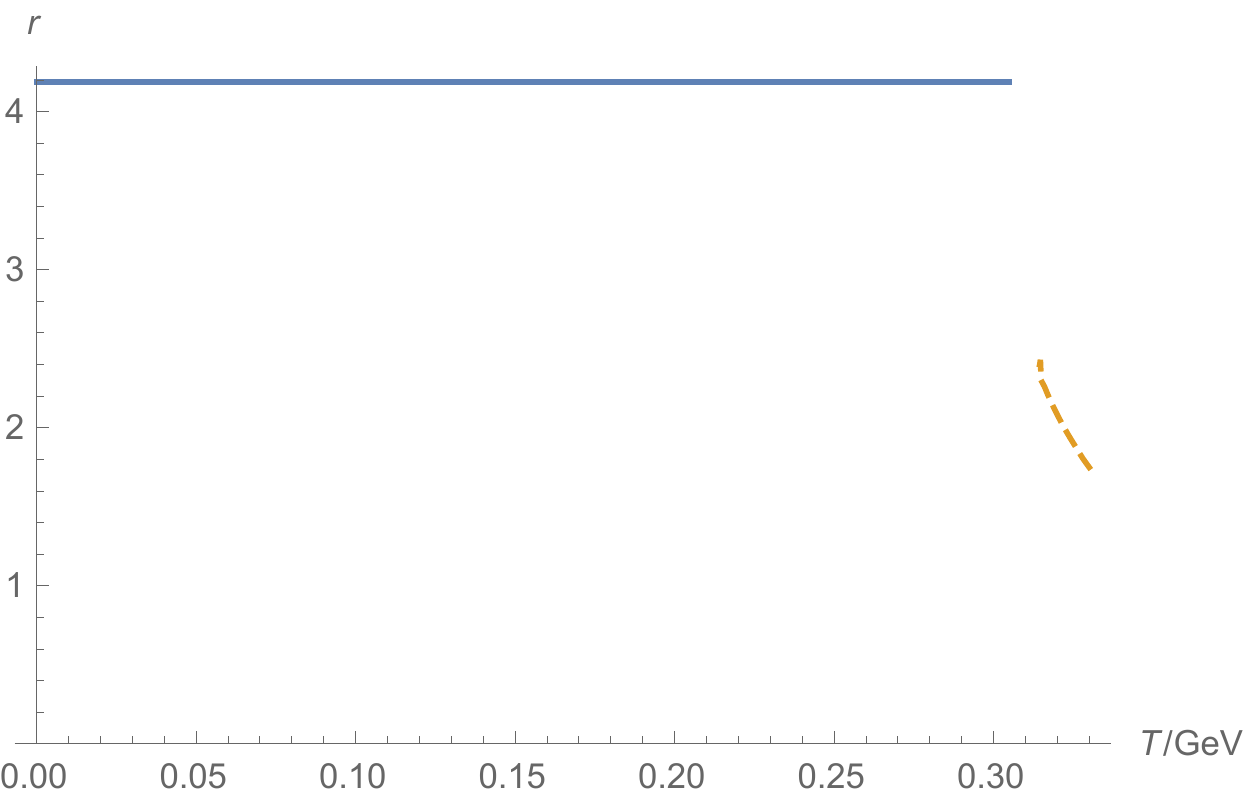} \caption{\label{susyopKRmanierr3}} \end{subfigure} \hfill \begin{subfigure}{0.35\textwidth} \includegraphics[width=\textwidth]{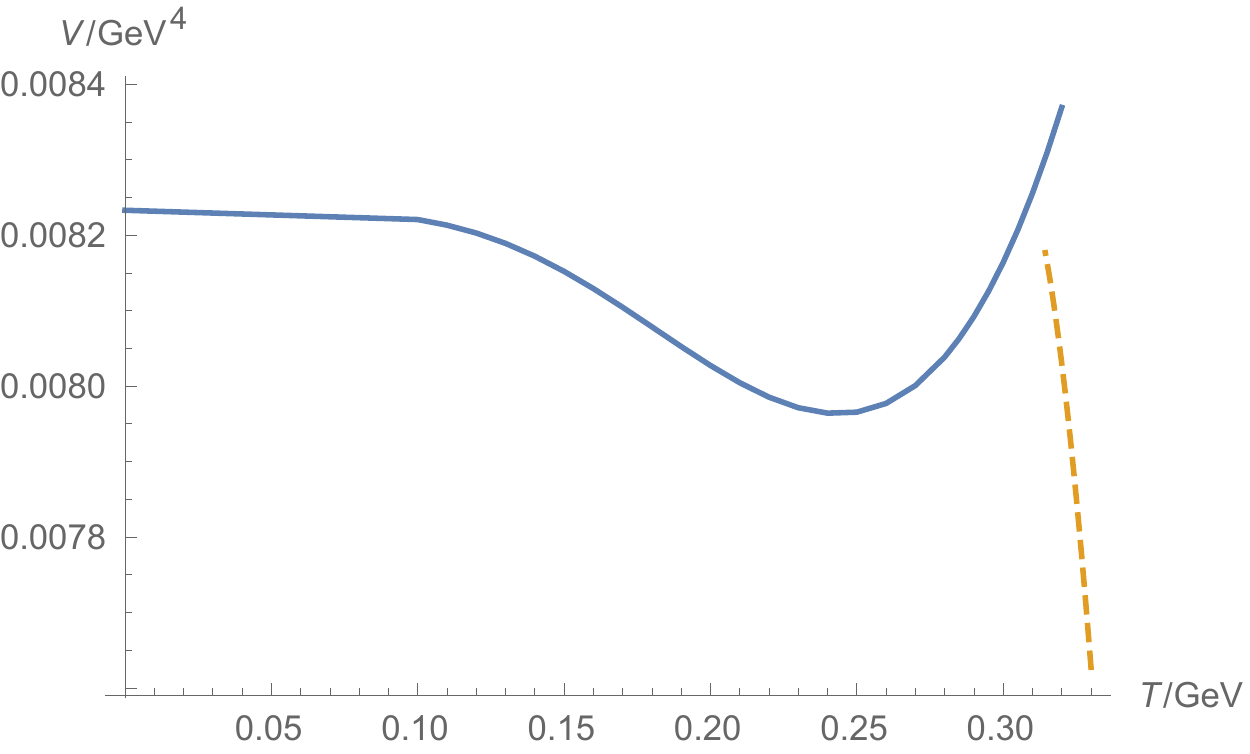} \caption{\label{susyopKRmanierV3}} \end{subfigure}
	\caption{Some of the results obtained in the KR approach for the symmetric case in SU(3) for the two minima we found: the confining minimum ($r = 4\pi/3 = 4.19$) in a full line and the deconfining minimum in a dashed line. The deconfining minimum is very shallow right above the transition temperature, making the numerics very unstable. (Minimization often ends up in the confining minimum.) This has resulted in a small gap in the data. Not shown are the dimension-two condensates, which do not vary much and also do not change much through the transition. \figurename\ \subref{susyopKRmanierlambda3} shows the Gribov parameter $\lambda$ and \figurename\ \subref{susyopKRmanierr2} shows the Polyakov loop $r$ as a function of temperature. \figurename\ \subref{susyopKRmanierV3} shows the vacuum energy. Extrapolating the vacuum energy of the deconfined minimum gives a first-order transition temperature at $T_c = \unit{0.310}{\giga\electronvolt}$. \label{susyopKRmanier3}}
\end{figure}

In the KR approach, the results are better. We find a second-order phase transition at $T_c = \unit{0.34}{\giga\electronvolt}$ for SU(2), see \figurename\ \ref{susyopKRmanierr2}. This is not too far from the lattice result in Table 6 in \cite{Lucini:2003zr}: $\unit{0.31}{\giga\electronvolt}$. For SU(3), we found the transition at $T_c = \unit{0.310}{\giga\electronvolt}$ (see \figurename\ \ref{susyopKRmanierr3}) and of first order (see \figurename\ \ref{susyopKRmanierr3}), again not too far from the lattice result of $\unit{0.28}{\giga\electronvolt}$ \cite{Lucini:2003zr}. The Gribov parameter $\lambda$ again goes up when going through the SU(3) transition, as seen in \figurename\ \ref{susyopKRmanierlambda3}.

The existence and orders of the transitions are in line with expectations, now. The transition temperatures are still on the high side, however. We tried playing with the scale parameter $\bar\mu$, but the results seem quite stable. We took $\bar\mu^2$ equal to the value of $m^2$ at zero temperature (for which the computations in \appendixname\ \ref{paramsymm} needed to be redone), which gave a smaller value of $\bar\mu^2$ and thus a higher value of the coupling constant $g^2$. We found a transition temperature of $T_c = \unit{0.35}{\giga\electronvolt}$ for SU(2): barely higher. With a higher coupling constant one could expect the finite-temperature corrections (which are all of first order in the coupling) to become more important, thus speeding up the transition. But changing $\bar\mu^2$ also modifies all the other zero-temperature parameters that enter the theory, and this seems to undo the effect.

\begin{figure}[t]
	\begin{subfigure}{0.35\textwidth} \includegraphics[width=\textwidth]{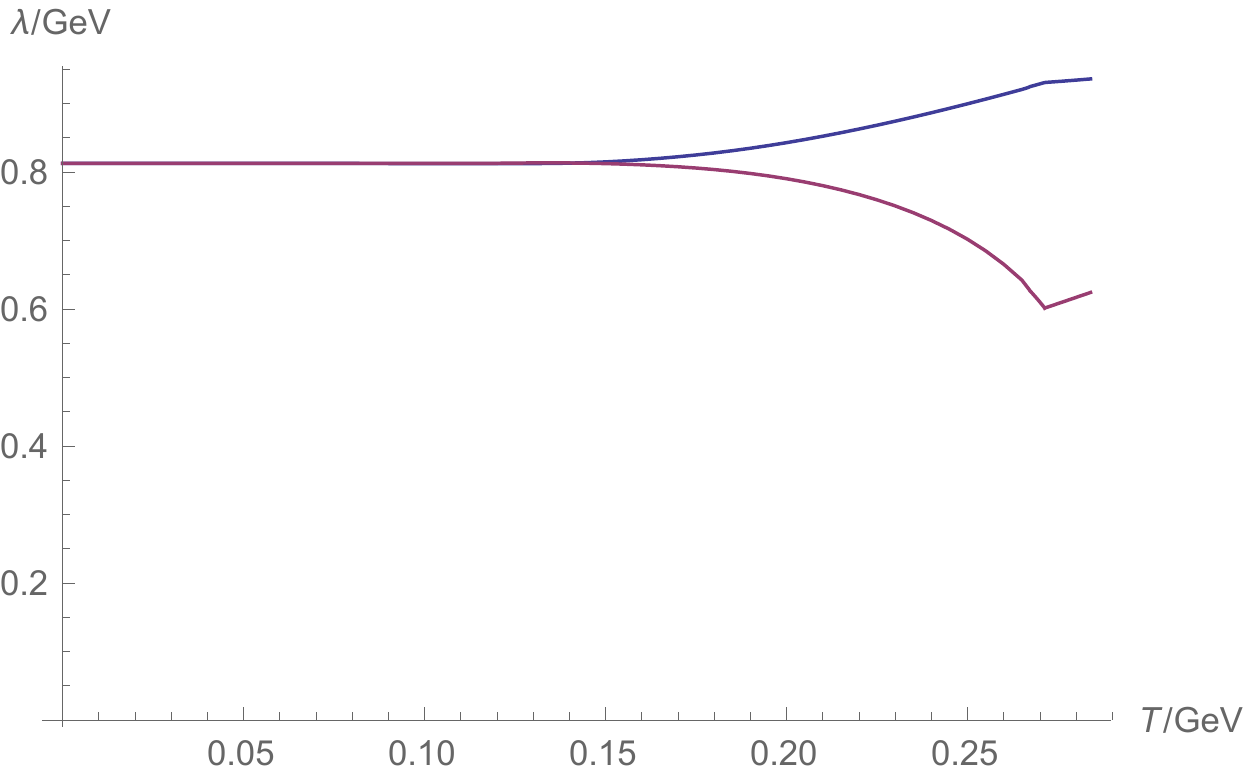} \caption{\label{susyopKRmaniernietontaardlambda2}} \end{subfigure} \qquad \begin{subfigure}{0.35\textwidth} \includegraphics[width=\textwidth]{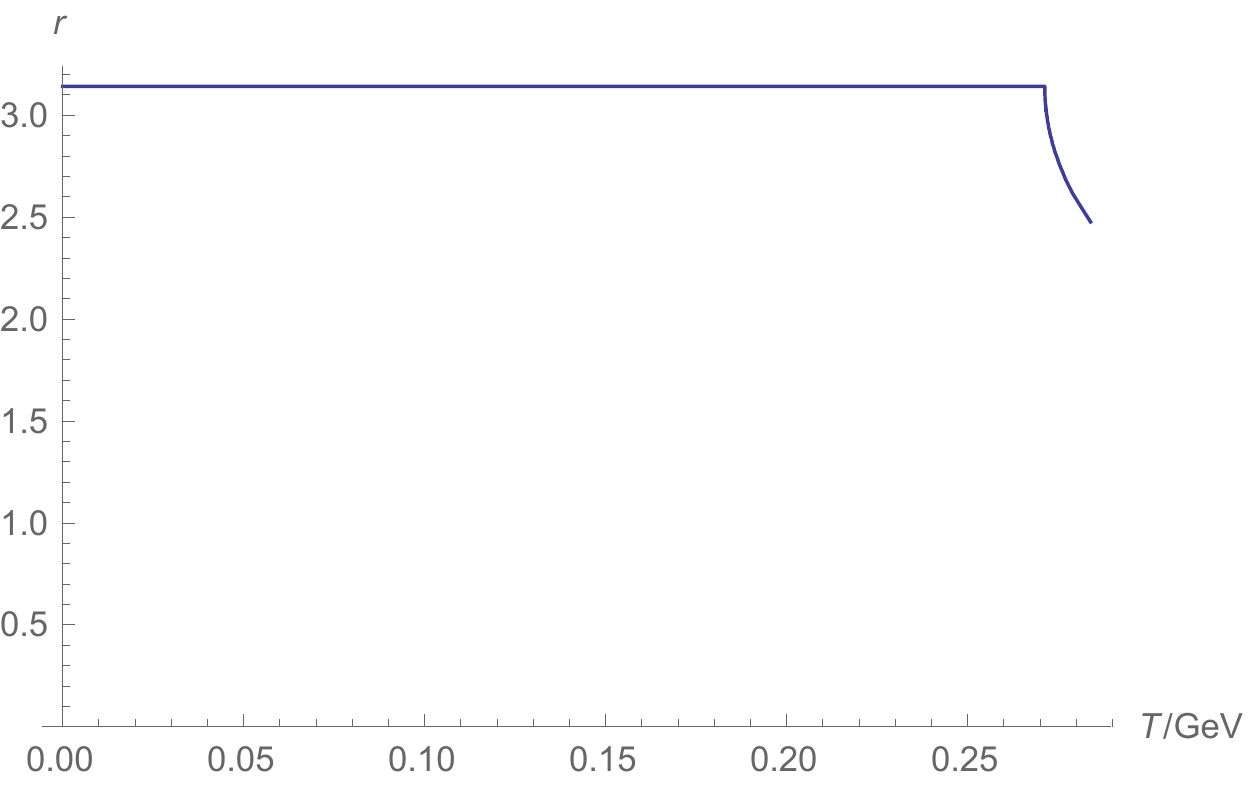} \caption{\label{susyopKRmaniernietontaardr2}} \end{subfigure}
	\caption{Some of the results obtained in the KR approach for the symmetric case in SU(2) for the partially degenerate approach to color-dependent Gribov parameters. Not shown are the dimension-two condensates, which do not vary much and also do not change much through the transition. \figurename\ \subref{susyopKRmaniernietontaardlambda2} shows the Gribov parameters $\lambda_0$ (upper line) and $\lambda_\text{ch}$ (lower line) and \figurename\ \subref{susyopKRmaniernietontaardr2} shows the Polyakov loop $r$ as a function of temperature. The second-order transition is now at $T_c = \unit{0.27}{\giga\electronvolt}$. \label{susyopKRmaniernietontaard2}}
\end{figure}

\begin{figure}[t]
	\begin{subfigure}{0.3\textwidth} \includegraphics[width=\textwidth]{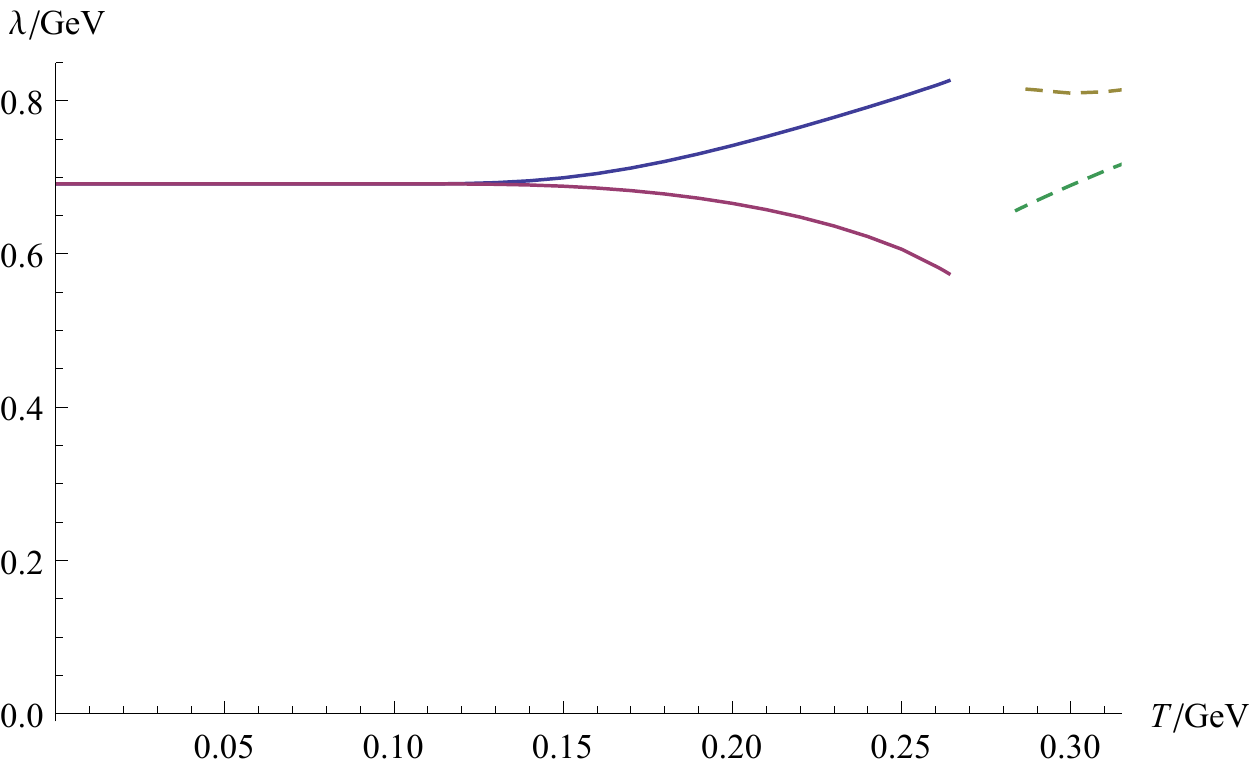} \caption{\label{susyopKRmaniernietontaardlambda3}} \end{subfigure} \hfill \begin{subfigure}{0.3\textwidth} \includegraphics[width=\textwidth]{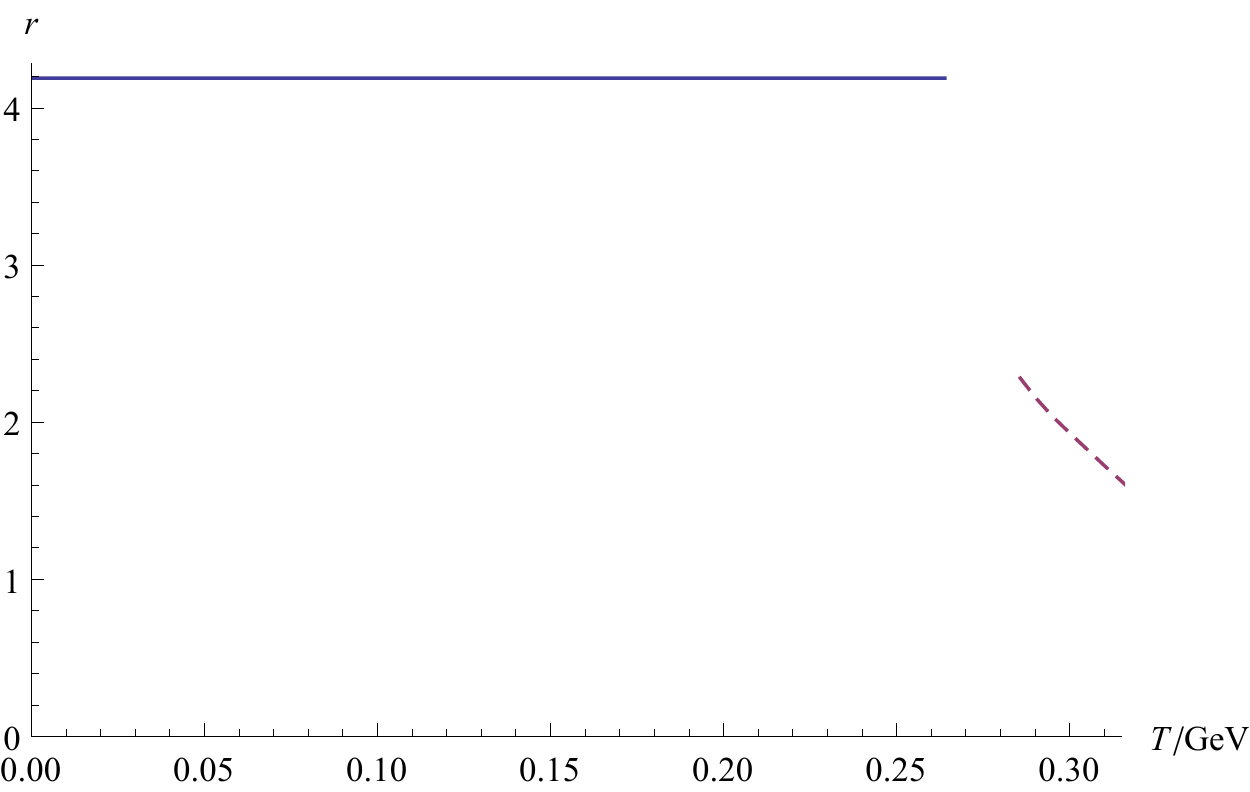} \caption{\label{susyopKRmaniernietontaardr3}} \end{subfigure} \hfill \begin{subfigure}{0.35\textwidth} \includegraphics[width=\textwidth]{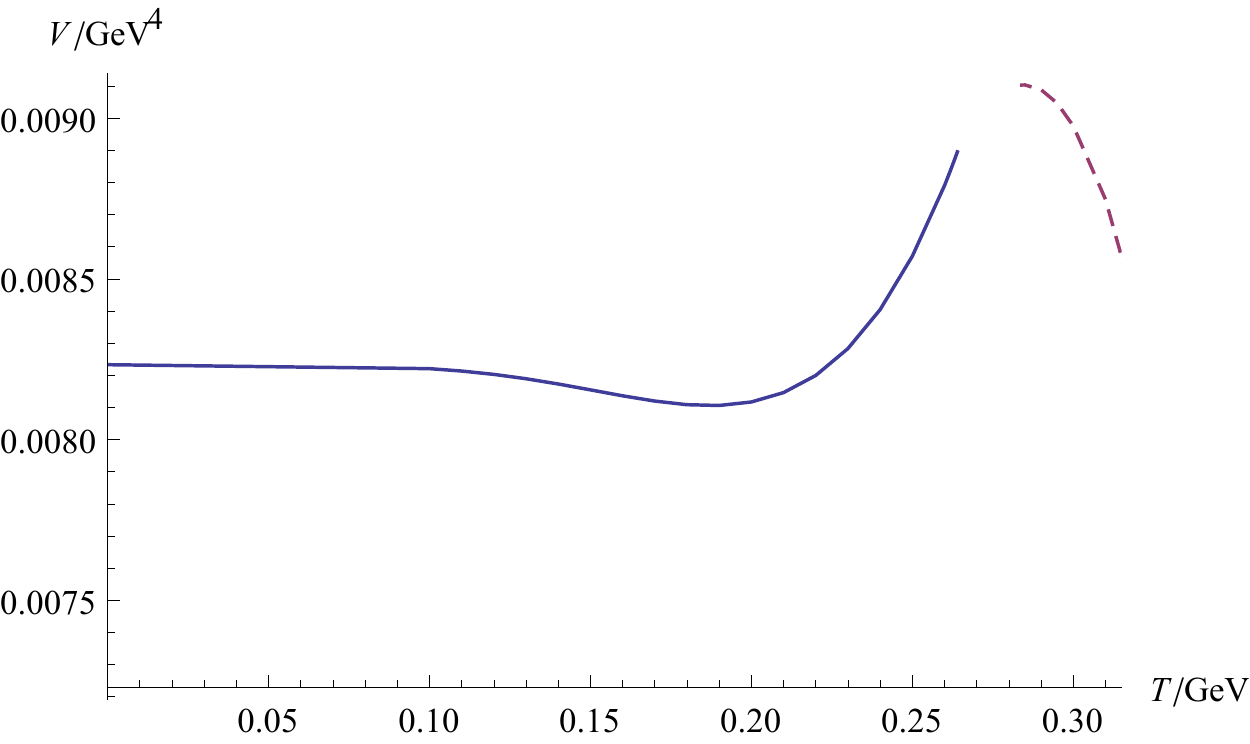} \caption{\label{susyopKRmaniernietontaardV3}} \end{subfigure}
	\caption{Some of the results obtained in the KR approach for the symmetric case in SU(3) for the partially degenerate approach to color-dependent Gribov parameters. Not shown are the dimension-two condensates, which do not vary much and also do not change much through the transition. We did not manage to find solutions between $T = \unit{0.264}{\giga\electronvolt}$ and $\unit{0.284}{\giga\electronvolt}$, due to the potential being nearly flat. As a result, we could only determine that the temperature of the (first-order) phase transition must be somewhere withing that range. \figurename\ \subref{susyopKRmaniernietontaardlambda3} shows the Gribov parameters $\lambda_0$ (upper line) and $\lambda_\text{ch}$ (lower line) and \figurename\ \subref{susyopKRmaniernietontaardr3} shows the Polyakov loop $r$ as a function of temperature. \figurename\ \subref{susyopKRmaniernietontaardV3} shows the value of the effective potential. \label{susyopKRmaniernietontaard3}}
\end{figure}

In \cite{Kroff:2018ncl}, Kroff and Reinosa also consider the introduction of different Gribov parameters in different color directions. In their paper, they find that doing so has a noteworthy impact on the transition temperature. We therefore also considered what they call the ``partially degenerate'' approach, where a ``neutral'' Gribov parameter $\gamma_0$ is coupled to the gluon fields in the Casimir and a ``charged'' one $\gamma_\text{ch}$ is coupled to the other modes. For SU(2) the transition temperature comes down with about a fifth to $T_c = \unit{0.27}{\giga\electronvolt}$ (see \figurename\ \ref{susyopKRmaniernietontaard2}), while for SU(3) the temperature of the (first-order) phase transition is between $T = \unit{0.264}{\giga\electronvolt}$ and $\unit{0.284}{\giga\electronvolt}$ (see \figurename\ \ref{susyopKRmaniernietontaard3}). Probably related to the flatness of the potential, we are unable to find a numerically more precise estimate of the transition temperature for SU(3).

\section{Conclusions}
In this paper we studied the Refined Gribov--Zwanziger action for SU(2) and SU(3) gauge theories with the Polyakov loop coupled to it
via the background field formalism. Doing so, we were able to compute the finite-temperature value of the Polyakov loop, the Gribov parameter, and the values of the dimension-two condensates simultaneously at the leading one-loop approximation.

We used several approaches. First there are two candidates for the Gribov auxiliary fields condensate that have been investigated in the past: $\langle\bar\varphi\varphi-\bar\omega\omega\rangle$ and $\langle\bar\varphi\varphi\rangle$ \cite{Dudal:2007cw,Dudal:2008sp,Dudal:2011gd}. The second one has enjoyed relatively more attention up to now, but from our results it turns out that only the first one (the more symmetric one) leads to phenomenologically acceptable results at finite temperature, where the second one does not. We furthermore used two different proposals to add a gluon background field to the Gribov formalism. The one proposed by the authors in \cite{Dudal:2017jfw} turns out to have issues, whereas the one proposed by Kroff and Reinosa \cite{Kroff:2018ncl} gives the best results.

From the point of view of physics, we found a second-order deconfinement phase transition for SU(2) and a first-order transition for SU(3), provided we used the symmetric condensate $\langle\bar\varphi\varphi-\bar\omega\omega\rangle$ and the Kroff--Reinosa approach. Just as in \cite{Canfora:2015yia}, the Gribov mass is nonzero at temperatures above $T_c$, indicating that the gluon propagator still violates positivity and as such it rather describes a quasiparticle than a ``free'' observable particle; see also \cite{Maas:2011se,Haas:2013hpa} for more on this.

Several improvements on the current setup can be proposed. First, one would expect the condensates to develop electric--magnetic asymmetries at finite temperature, in the vein of \cite{Chernodub:2008kf,Bornyakov:2016geh}. This markedly complicates the computations, and previous work has shown that the results are not greatly impacted \cite{Dudal:2022nnu}. Another possibility is, naturally, to go to two-loop order. The Kroff--Reinosa approach is computationally the most elegant and simplest one, and luckily it turned out to be the best one phenomenologically as well. This allows one to hope that a two-loop computation would be tractable, although the two-loop generalization of \cite{Kroff:2018ncl} without any extra condensates is also still lacking. It would also be interesting to test in practice the argument in \cite{Kroff:2018ncl} that the KR model is renormalizable to two-loop order as well. A full BRST based analysis of this feature to all orders looks too ambitious given the presence of the non-local dressing factors as in \eqref{WilsonKR}. Furthermore, it remains an open question how to split the path integration over the various twisted sectors when the auxiliary Stueckelberg-like $h$-field is brought on-shell.

\section*{Acknowledgements}
D.V. is grateful for the hospitality at KU Leuven, where parts of this work were done, made possible through the KU Leuven Senior Fellowship SF/19/008 and IF project C14/21/087. We would also like to thank Diego R. Granado for the check of the SU(3) calculus.

\appendix
\section{Proof of sufficiency of Gribov construction applied to $-\bar D^h_\mu D^h_\mu$} \label{DDproof}
In this section, we prove that the modification of the Gribov--Zwanziger action as given in \eqref{s0ldwh} is sufficient to remove infinitesimal Gribov copies in the Landau--DeWitt gauge with background $\bar A_\mu$.

The Faddeev--Popov operator in the Landau background gauge is $-\bar D_\mu^{ab} D_\mu^{bc}$. As shown in \cite{Dudal:2017jfw}, however, basing the Gribov construction on this operator leads to a breaking of background gauge symmetry $\delta \bar A_\mu^a = \bar D_\mu^{ab} \beta^b$ with $\beta^a$ the gauge parameter. In \cite{Dudal:2017jfw}, the operator $-\partial_\mu (D^h)_\mu^{ab}$ was proposed. In the case at hand, however, we have a nonzero $(\bar A^h)_\mu^a$, such that we need to use $-(\bar D^h)_\mu^{ab} (D^h)_\mu^{bc}$. (In this operator, the first covariant derivative contains the transformed background $(\bar A^h)_\mu^a$, the second one contains the full field $(a^h)_\mu^a$.)

Let us now prove that this is correct. To do so, let us use a shorthand notation from here on to avoid clutter of colour and Lorentz indices, writing $-\bar D^h D^h$ and $D=\partial+a$ etc. We want to prove that restricting the path integral to configurations with $-\bar D^h D^h>0$ actually excludes (almost) all Gribov copies related to the zero modes of the Faddeev--Popov operator $-\bar D D$. Given a configuration in the permissible space $-\bar D^h D^h>0$, assume the exists a zero mode $\xi$ of $-\bar D D$:
\begin{equation}
	-\bar D D \xi = 0 \;.
\end{equation}
To prove that this implies $\xi=0$, we will assume $\xi$ can be written as a series in the background $\xi = \sum_{n=0}^\infty \bar A^n \xi_n[\mathcal A]$. We can rewrite the equation for $\xi$ as
\begin{equation}
	-\bar D^h D^h \xi + \bar A^h D^h \xi - \bar A D \xi + \partial ((a^h-a)\xi) = 0 \;.
\end{equation}
Due to the assumed invertibility of $-\bar D^h D^h$, this means that
\begin{equation} \label{xieq}
	\xi = \frac1{-\bar D^h D^h } \Big( -\bar A^h D^h \xi + \bar A D \xi - \partial ((a^h-a)\xi) \Big) \;.
\end{equation}
In the limit $\bar A\to 0$, we have that $\bar A^h\to0$, such that $\bar A^h = \mathcal O(\bar A)$. Furthermore in the same limit the gauge condition for $a^h$ becomes identical to that for $a$, such that also $a^h-a\to0$. This means that the right-hand side of \eqref{xieq} starts at at least one order in $\bar A$ higher than $\xi$, which can never be equal to $\xi$ except if $\xi=0$. This concludes the proof that restricting the path integral to configurations with $-\bar D^h D^h>0$ actually excludes all Gribov copies related to the zero modes of the Faddeev--Popov operator $-\bar D D$ that are expressable as a Taylor series in the background field, \emph{i.e.}~that are continuous deformations around the zero background (standard Landau gauge).

This completes our proof.

\section{The projection operator in equation \eqref{cdGp}} \label{KRproj}
We want to construct (in the notations of \cite{Kroff:2018ncl}, see equation (26)) a background-gauge invariant equivalent to $\gamma_\kappa^2 f^{\kappa\lambda\eta} A_\mu^\kappa (\varphi_\mu^{\lambda\eta} + \bar\varphi_\mu^{\lambda\eta})$. Under background gauge transformations, one has
\begin{subequations} \begin{gather}
	\delta \bar A_\mu^a = \bar D_\mu^{ab} \varpi^b \;, \\
	\delta A_\mu^a = -gf^{abc}\varpi^bA_\mu^c \;, \label{bgmatter}
\end{gather} \end{subequations}
and transformations analogous to \eqref{bgmatter} for $\varphi$ and $\bar\varphi$. In \cite{Kroff:2018ncl} the authors state that this is possible, but without showing explicitly how. If the background is a constant and the transformation brings it to another constant background (for example a gauge rotation) then the expression show in equation (26) in \cite{Kroff:2018ncl} is manifestly invariant provided we remember to redefine the indices. (The Greek color indices in \cite{Kroff:2018ncl} are defined with respect to the Casimir, where the background is assumed to be in.) To get invariance under general background transformations, we need to do more work.

We need to define a projection operator $P^{ab}$ such that
\begin{equation}
	f^{acd} P^{ab} A_\mu^b (\varphi_\mu^{cd} + \bar\varphi_\mu^{cd})
\end{equation}
is invariant. If the background is in the minimal Landau gauge, we want this projection operator to be equal to $P^{ab} \to \bar A_\mu^a\bar A_\mu^b/\bar A^2$. In that case, the projector will pick out the color direction along the background, to which we couple one of the $\gamma_0$'s. For example in SU(2) there is only one Casimir direction and we can therefore use
\begin{equation}
	\Big(\gamma_0P^{ab} + \gamma_{\text{ch}} (\delta^{ab}-P^{ab})\Big) A_\mu^b \;,
\end{equation}
In order to write down such a projector, we search for a field $\bar{\mathcal A}_\mu^a$ such that $\bar{\mathcal A}_\mu^a$ transforms as \eqref{bgmatter} under background transformations ($\delta \bar{\mathcal A}_\mu^a = -gf^{abc} \varpi^b \bar{\mathcal A}_\mu^c$) and also such that $\bar{\mathcal A}_\mu^a \to \bar A_\mu^a$ whenever the background is in minimal Landau gauge. Then,
\begin{equation}
	P^{ab} = \frac{\bar{\mathcal A}_\mu^a\bar{\mathcal A}_\mu^b}{\bar{\mathcal A}^2}
\end{equation}
fits the bill:
\begin{equation}
	\delta P^{ab} = - g \frac{f^{acd}\varpi^c\bar{\mathcal A}_\mu^d\bar{\mathcal A}_\mu^b + f^{bcd} \bar{\mathcal A}_\mu^a\varpi^c \bar{\mathcal A}_\mu^d}{\bar{\mathcal A}^2} = - g (f^{acd}\delta^{be}+\delta^{ad}f^{bce}) \varpi^c P^{de} \quad
	\Rightarrow\quad \delta(P^{ab} A_\mu^b) = -gf^{abc}\varpi^bP^{cd}A_\mu^d \;,
\end{equation}
which is sufficient for our needs.

Take the Ansatz
\begin{subequations} \begin{equation}
	\bar{\mathcal A}_\mu^a = (\bar A^h)_\mu^a + X_\mu^a \;.
\end{equation}
Expand the above in orders of $\bar A_\mu^a$:
\begin{gather}
	(\bar A^h)_\mu^a = \left(\delta_{\mu\nu} - \tfrac{\partial_\mu\partial_\nu}{\partial^2}\right) \sum_{n=1}^\infty (\mathcal F_n)_\nu^a(\bar A) \;, \\
	X_\mu^a = \sum_{n=2}^\infty (\mathcal G_n)_\mu^a(\bar A) \;,
\end{gather} \end{subequations}
where the index $n$ denotes the number of $\bar A_\mu^a$ fields. Given that $(\bar A^h)_\mu^a$ is invariant under $\delta \bar A_\mu^a = \bar D_\mu^{ab} \varpi^b$, we get
\begin{equation}
	\delta \bar{\mathcal A}_\mu^a = \delta X_\mu^a = \sum_{n=2}^\infty \bar D_\nu^{bc} \varpi^c \frac{\delta(\mathcal G_n)_\mu^a}{\delta \bar A_\nu^b}(\bar A) = \sum_{n=1}^\infty \partial_\nu \varpi^b \frac{\delta(\mathcal G_{n+1})_\mu^a}{\delta \bar A_\nu^b}(\bar A) - gf^{bcd} \bar A_\nu^d \varpi^c \sum_{n=2}^\infty \frac{\delta(\mathcal G_n)_\mu^a}{\delta \bar A_\nu^b}(\bar A) \;.
\end{equation}
Requiring $\delta \bar{\mathcal A}_\mu^a = -gf^{abc} \varpi^b \bar{\mathcal A}_\mu^c$ and equating order by order in $\bar A_\mu^a$ gives
\begin{multline}
	-gf^{abc} \varpi^b \left(\delta_{\mu\nu} - \tfrac{\partial_\mu\partial_\nu}{\partial^2}\right) \sum_{n=1}^\infty (\mathcal F_n)_\nu^c(\bar A) - gf^{abc} \varpi^b \sum_{n=2}^\infty (\mathcal G_n)_\mu^c(\bar A) \\ = \sum_{n=1}^\infty \partial_\nu \varpi^b \frac{\delta(\mathcal G_{n+1})_\mu^a}{\delta \bar A_\nu^b}(\bar A) - gf^{bcd} \bar A_\nu^d \varpi^c \sum_{n=2}^\infty \frac{\delta(\mathcal G_n)_\mu^a}{\delta \bar A_\nu^b}(\bar A) \\
	\Rightarrow \quad \partial_\nu \varpi^b \frac{\delta(\mathcal G_{n+1})_\mu^a}{\delta \bar A_\nu^b}(\bar A) = - gf^{abc} \varpi^b \left(\delta_{\mu\nu} - \tfrac{\partial_\mu\partial_\nu}{\partial^2}\right) (\mathcal F_n)_\nu^c(\bar A) \\ - gf^{abc} \varpi^b (\mathcal G_n)_\mu^c(\bar A) + gf^{bcd} \bar A_\nu^d \varpi^c \frac{\delta(\mathcal G_n)_\mu^a}{\delta \bar A_\nu^b}(\bar A) \;.
\end{multline}

For $n=1$ one gets (with $\mathcal G_1=0$):
\begin{equation}
	\partial_\nu \varpi^b \frac{\delta(\mathcal G_2)_\mu^a}{\delta \bar A_\nu^b}(\bar A) = - gf^{abc} \varpi^b \left(\delta_{\mu\nu} - \tfrac{\partial_\mu\partial_\nu}{\partial^2}\right) \bar A_\nu^c
\end{equation}
Given that
\begin{equation}
	\partial_\nu \omega^b \frac\delta{\delta \bar A_\nu^b} \left(\delta_{\mu\nu} - \tfrac{\partial_\mu\partial_\nu}{\partial^2}\right) \bar A_\nu^a = \left(\delta_{\mu\nu} - \tfrac{\partial_\mu\partial_\nu}{\partial^2}\right) \partial_\nu \varpi^a = 0 \;,
\end{equation}
we only need to multiply $\left(\delta_{\mu\nu} - \tfrac{\partial_\mu\partial_\nu}{\partial^2}\right) \bar A_\nu^b$ with some expression $Y^{ab}$ that obeys $\partial_\nu \varpi^b \frac\delta{\delta \bar A_\nu^b} Y^{ac} = - gf^{abc} \varpi^b$. An obvious solutions is $Y^{bc} = g f^{abc} \tfrac{\partial_\lambda}{\partial^2} \bar A_\lambda^c$.

The cases for $n>1$ are left as an exercise for the reader. The final result is (to second order in the background field):
\begin{equation}
	\bar{\mathcal A}_\mu^a = \left(\delta_{\mu\nu} - \tfrac{\partial_\mu\partial_\nu}{\partial^2}\right) \left(\bar A_\nu^a - g f^{abc} \left((\delta_{\nu\lambda} - \tfrac12 \tfrac{\partial_\nu\partial_\lambda}{\partial^2}) \bar A_\lambda^b\right) \tfrac{\partial_\kappa}{\partial^2} \bar A_\kappa^c + \cdots\right) + g f^{abc} \left((\delta_{\mu\nu} - \tfrac{\partial_\mu\partial_\nu}{\partial^2}) \bar A_\nu^b\right) \tfrac{\partial_\lambda}{\partial^2} \bar A_\lambda^c + \cdots
\end{equation}

\section{Free parameters in the symmetric approach} \label{paramsymm}
The free parameters of the $\bar\varphi\varphi$ approach at zero temperature were computed in \cite{Dudal:2019ing}. The symmetric approach has not yet been done, so we work it out in this \appendixname.

The gap equation for $\lambda^4$ is
\begin{equation}
	\frac56 - \frac43 b_0 - \frac{4(4\pi)^2}{3Ng^2} - \frac12 \log\frac{m^2M^2 + \lambda^4}{\bar\mu^4} - \frac{m^2+M^2}{\sqrt{4\lambda^4-(m^2-M^2)^2}}\operatorname{arccot}\frac{m^2+M^2}{\sqrt{4\lambda^4-(m^2-M^2)^2}} = 0 \;,
\end{equation}
or, after plugging in the renormalization group, the poles, and our choice for $\bar\mu^2$:
\begin{equation}
	\frac56 - \frac43 b_0 - \frac{44}9 \ln\frac{\sqrt{x_0^2+y_0^2}}{\Lambda_{\overline{\text{MS}}}^2} \\ - \frac{x_0}{y_0}\operatorname{arccot}\frac{x_0}{y_0} = 0 \;.
\end{equation}
This gives $b_0 = -8.49$ in SU(3) and $-6.7$ in SU(2).

The equation for $m^2$ is
\begin{multline}
	\frac{2(m^2M^2 + \lambda^4)}{\sqrt{4\lambda^4-(m^2-M^2)^2}} \operatorname{arccot}\frac{m^2+M^2}{\sqrt{4\lambda^4-(m^2-M^2)^2}} \\ + m^2 \left(\frac13 - \frac6{13} \frac{(4\pi)^2}{g^2N} - \frac{m^2+M^2}{\sqrt{4\lambda^4-(m^2-M^2)^2}} \operatorname{arccot}\frac{m^2+M^2}{\sqrt{4\lambda^4-(m^2-M^2)^2}} - \frac12 \ln\frac{m^2M^2 + \lambda^4}{\bar\mu^4}\right) = 0 \;,
\end{multline}
or, after plugging in the renormalization group, the poles, and our choice for $\bar\mu^2$:
\begin{equation}
	\frac{x_0^2+y_0^2}{y_0} \operatorname{arccot}\frac{x_0}{y_0} \\ + m^2 \left(\frac13 - \frac{22}{13} \ln\frac{\sqrt{x_0^2+y_0^2}}{\Lambda_{\overline{\text{MS}}}^2} - \frac{x_0}{y_0} \operatorname{arccot}\frac{x_0}{y_0}\right) = 0 \;.
\end{equation}
This gives $m^2 = \unit{0.152}{\giga\electronvolt^2}$ in SU(3) and $\unit{0.27}{\giga\electronvolt^2}$ in SU(2). Given that $M^2 = 2x_0-m^2$, we also find $M^2 = \unit{0.370}{\giga\electronvolt^2}$ in SU(3) and $\unit{0.31}{\giga\electronvolt^2}$ in SU(2). Given that $\lambda^4=x_0^2+y_0^2-m^2M^2$ we also find $\lambda^2 = \unit{0.478}{\giga\electronvolt^2}$ in SU(3) and $\unit{0.67}{\giga\electronvolt^2}$ in SU(2).

The equation for $M^2$ is
\begin{multline}
	\frac{2(m^2M^2 + \lambda^4)}{\sqrt{4\lambda^4-(m^2-M^2)^2}} \operatorname{arccot}\frac{m^2+M^2}{\sqrt{4\lambda^4-(m^2-M^2)^2}} \\ + M^2 \left(-\frac23 \frac{(4\pi)^2}{N^2-1} \beta - \frac{m^2+M^2}{\sqrt{4\lambda^4-(m^2-M^2)^2}} \operatorname{arccot}\frac{m^2+M^2}{\sqrt{4\lambda^4-(m^2-M^2)^2}} - \frac12 \ln\frac{m^2M^2 + \lambda^4}{M^4}\right) = 0 \;,
\end{multline}
or, after plugging in the poles:
\begin{equation}
	\frac{x_0^2+y_0^2}{y_0} \operatorname{arccot}\frac{x_0}{y_0} \\ + M^2 \left(-\frac23 \frac{(4\pi)^2}{N^2-1} \beta - \frac{x_0}{y_0} \operatorname{arccot}\frac{x_0}{y_0} - \frac12 \ln\frac{x_0^2+y_0^2}{M^4}\right) = 0 \;.
\end{equation}
This gives $\beta = 0.0601$ in SU(3) and $0.045$ in SU(2).

\section{Conventions}
\subsection{SU(2)} \label{sutwoapp}
We define isospin eigenstates as
\begin{equation} \label{isospin}
	\mx v_+ = \frac1{\sqrt2} \begin{pmatrix} i \\ 1 \\ 0 \end{pmatrix} \;, \qquad \mx v_- = \frac1{\sqrt2} \begin{pmatrix} i \\ -1 \\ 0 \end{pmatrix} \;, \qquad \mx v_0 = \begin{pmatrix} 0 \\ 0 \\ 1 \end{pmatrix} \;.
\end{equation}
We then have that
\begin{subequations} \begin{gather}
	\mathbbm 1 = \mx v_+ \mx v_+^\dagger + \mx v_- \mx v_-^\dagger + \mx v_3 \mx v_3^\dagger \label{spinunity} \\
	\tr \mx A = \mx v_+^\dagger \mx A \mx v_+ + \mx v_-^\dagger \mx A \mx v_- + \mx v_3^\dagger \mx A \mx v_3 \;.
\end{gather} \end{subequations}
If we define
\begin{equation}
	\mx s^{ab} = i\epsilon^{ab3} \;, \qquad \mx a_+^{ab} = \epsilon^{ab1} - i\epsilon^{ab2} \;, \qquad \mx a_-^{ab} = -\epsilon^{ab1} - i\epsilon^{ab2} \;.
\end{equation}
we have the commutation relations
\begin{equation}
	[\mx s,\mx a_\pm] = \pm\mx a_\pm \;, \qquad [\mx a_+,\mx a_-] = 2\mx s \;,
\end{equation}
and that
\begin{subequations} \begin{gather}
	\mx s \mx v_3 = 0 \;, \qquad \mx s \mx v_s = s \mx v_s \;, \\
	\mx a_+ \mx v_+ = 0 \;, \qquad \mx a_+ \mx v_0 = \sqrt2 \mx v_+ \;, \qquad \mx a_+ \mx v_- = \sqrt2 \mx v_0 \;, \\
	\mx a_- \mx v_+ = \sqrt2 \mx v_0 \;, \qquad \mx a_- \mx v_0 = \sqrt2 \mx v_- \;, \qquad \mx a_- \mx v_- = 0 \;.
\end{gather} \end{subequations}
We also have
\begin{equation} \label{epsilons}
	\epsilon^{ace} \mathcal O^{cd} \epsilon^{bde} = \left(\mx s \mathcal O \mx s + \tfrac12 \mx a_+ \mathcal O \mx a_- + \tfrac12 \mx a_- \mathcal O \mx a_+\right)^{ab} \;.
\end{equation}

\subsection{SU(3)} \label{suthreeapp}
The structure constants of the Lie algebra of SU(3) are given by
\begin{subequations} \begin{gather}
	f_{123} = 1 \;, \\
	f_{147} = -f_{156} = f_{246} = f_{257} = f_{345} = -f_{367} = \frac12 \;, \\
	f_{458} = f_{678} = \frac{\sqrt3}2 \;,
\end{gather} \end{subequations}
while all other $f_{abc}$ not related to these by permutation are zero. To avoid cluttered indices, define the matrices $(f^a)_{bc} = f_{abc}$. Now define the following operators:
\begin{subequations} \begin{gather}
	\mx s_3 = if^3 \;, \qquad \mx s_8 = if^8 \;, \\
	\mx a_1^\pm = \pm f^1 - i f^2 \;, \qquad \mx a_2^\pm = \pm f^4 - i f^5 \;, \qquad \mx a_3^\pm = \pm f^6 - i f^7 \;.
\end{gather} \end{subequations}
These obey the commutation relations:
\begin{subequations} \begin{gather}
	[\mx s_3,\mx s_8] = 0 \;, \\
	[\mx s_3,\mx a_1^\pm] = \pm \mx a_1^\pm \;, \qquad [\mx s_8,\mx a_1^\pm] = 0 \;, \\
	[\mx s_3,\mx a_2^\pm] = \pm \tfrac 12 \mx a_2^\pm \;, \qquad [\mx s_8,\mx a_2^\pm] = \pm \tfrac{\sqrt3}2 \mx a_2^\pm \;, \\
	[\mx s_3,\mx a_3^\pm] = \mp \tfrac 12 \mx a_3^\pm \;, \qquad [\mx s_8,\mx a_3^\pm] = \pm \tfrac{\sqrt3}2 \mx a_3^\pm \;, \\
	[\mx a_1^+,\mx a_1^-] = 2\mx s_3 \;, \qquad [\mx a_2^+,\mx a_2^-] = \mx s_3 + \sqrt3 \mx s_8 \;, \qquad [\mx a_3^+,\mx a_3^-] = -\mx s_3 + \sqrt3 \mx s_8 \;, \\
	[\mx a_1^\pm,\mx a_2^\pm] = 0 \;, \qquad [\mx a_1^\pm,\mx a_2^\mp] = -i \mx a_3^\mp \;, \\
	[\mx a_1^\pm,\mx a_3^\pm] = -i \mx a_2^\pm \;, \qquad [\mx a_1^\pm,\mx a_3^\mp] = 0 \;, \\
	[\mx a_2^\pm,\mx a_3^\pm] = 0 \;, \qquad [\mx a_2^\pm,\mx a_3^\mp] = -i \mx a_1^\pm \;.
\end{gather} \end{subequations}

Next, define the following vectors:
\begin{subequations} \begin{gather}
	 \mx v_3^a = \delta^{a3} \;, \qquad \mx v_8^a = \delta^{a8} \;, \\
	(\mx v_1^\pm)^a = \frac1{\sqrt2} (i\delta^{a1} \pm \delta^{a2}) \;, \qquad (\mx v_2^\pm)^a = \frac1{\sqrt2} (i\delta^{a4} \pm \delta^{a5}) \;, \qquad (\mx v_3^\pm)^a = \frac1{\sqrt2} (i\delta^{a6} \pm \delta^{a7}) \;.
\end{gather} \end{subequations}
We have the following operations: \\
\hspace*{1cm}\begin{tabular}{l*{8}{|c}}
 & $\mx s_3$ & $\mx s_8$ & $\mx a_1^+$ & $\mx a_1^-$ & $\mx a_2^+$ & $\mx a_2^-$ & $\mx a_3^+$ & $\mx a_3^-$ \\ \hline
$\mx v_3$ & 0 & 0 & $\sqrt2\mx v_1^+$ & $\sqrt2\mx v_1^-$ & $\tfrac1{\sqrt2}\mx v_2^+$ & $\tfrac1{\sqrt2}\mx v_2^-$ & $-\tfrac1{\sqrt2}\mx v_3^+$ & $-\tfrac1{\sqrt2}\mx v_3^-$ \\ \hline
$\mx v_8$ & 0 & 0 & 0 & 0 & $\sqrt{\tfrac32}\mx v_2^+$ & $\sqrt{\tfrac32}\mx v_2^-$ & $\sqrt{\tfrac32}\mx v_3^+$ & $\sqrt{\tfrac32}\mx v_3^-$ \\ \hline
$\mx v_1^+$ & $\mx v_1^+$ & 0 & 0 & $\sqrt2\mx v_3$ & 0 & $i\mx v_3^-$ & $i\mx v_2^+$ & 0 \\ \hline
$\mx v_1^-$ & $-\mx v_1^-$ & 0 & $\sqrt2\mx v_3$ & 0 & $i\mx v_3^+$ & 0 & 0 & $i\mx v_2^-$ \\ \hline
$\mx v_2^+$ & $\tfrac12\mx v_2^+$ & $\tfrac{\sqrt3}2\mx v_2^+$ & 0 & $i\mx v_3^+$ & 0 & $\tfrac1{\sqrt2}(\mx v_3+{\sqrt3}\mx v_8)$ & 0 & $-i\mx v_1^+$ \\ \hline
$\mx v_2^-$ & $-\tfrac12\mx v_2^-$ & $-\tfrac{\sqrt3}2\mx v_2^-$ & $i\mx v_3^-$ & 0 & $\tfrac1{\sqrt2}(\mx v_3+{\sqrt3}\mx v_8)$ & 0 & $-i\mx v_1^-$ & 0 \\ \hline
$\mx v_3^+$ & $-\tfrac12\mx v_3^+$ & $\tfrac{\sqrt3}2\mx v_3^+$ & $-i\mx v_2^+$ & 0 & 0 & $-i\mx v_1^-$ & 0 & $\tfrac1{\sqrt2}(-\mx v_3+{\sqrt3}\mx v_8)$ \\ \hline
$\mx v_3^-$ & $\tfrac12\mx v_3^-$ & $-\tfrac{\sqrt3}2\mx v_3^-$ & 0 & $-i\mx v_2^-$ & $-i\mx v_1^+$ & 0 & $\tfrac1{\sqrt2}(-\mx v_3+{\sqrt3}\mx v_8)$ & 0
\end{tabular}
\ \\As a result, the $\mx a$'s function as ladder operators:
\begin{subequations} \begin{gather}
	\mx a_1^\pm\;: \quad 0\leftarrow\mx v_1^-\leftrightarrow\mx v_3\leftrightarrow\mx v_1^+\rightarrow0 \;, \quad 0\leftarrow\mx v_2^-\leftrightarrow\mx v_3^-\rightarrow0 \;, \quad 0\leftarrow\mx v_3^+\leftrightarrow\mx v_2^+\rightarrow0 \;, \quad 0\leftarrow\mx v_8\rightarrow0 \;, \\
	\mx a_2^\pm\;: \quad 0\leftarrow\mx v_2^-\leftrightarrow(\mx v_3,\mx v_8)\leftrightarrow\mx v_2^+\rightarrow0 \;, \quad 0\leftarrow\mx v_1^-\leftrightarrow\mx v_3^+\rightarrow0 \;, \quad 0\leftarrow\mx v_3^-\leftrightarrow\mx v_1^+\rightarrow0 \;, \\
	\mx a_3^\pm\;: \quad 0\leftarrow\mx v_3^-\leftrightarrow(\mx v_3,\mx v_8)\leftrightarrow\mx v_3^+\rightarrow0 \;, \quad 0\leftarrow\mx v_2^-\leftrightarrow\mx v_1^-\rightarrow0 \;, \quad 0\leftarrow\mx v_1^+\leftrightarrow\mx v_2^+\rightarrow0 \;,
\end{gather} \end{subequations}
where the plus operators work to the right and the minus operators to the left.

Now consider the operator $f_{ace}\mathcal O^{cd} f_{dbe}$. In the above notations, this gives
\begin{equation} \label{operatorfof}
	\left(\mx s_3\mathcal O\mx s_3 + \mx s_8\mathcal O\mx s_8 + \tfrac12 \mx a_i^+\mathcal O\mx a_i^- + \tfrac12 \mx a_i^-\mathcal O\mx a_i^+\right)_{ab} \;.
\end{equation}
Assuming $\mathcal O^{ab}$ to be diagonal in the above basis, the operator under consideration is also diagonal in the $\mx v_i^\pm$ subspace with eigenvalues
\begin{subequations} \begin{align}
\label{eigenvaluesv1}
	\mx v_1^\pm \;: & \quad \mathcal O_3 + \mathcal O_1^\pm + \frac12\mathcal O_2^\pm + \frac12\mathcal O_3^\mp \;, \\
\label{eigenvaluesv2}
	\mx v_2^\pm \;: & \quad \frac14\mathcal O_3 + \frac34\mathcal O_8 + \frac12\mathcal O_1^\pm + \mathcal O_2^\pm + \frac12\mathcal O_3^\pm \;, \\
	\mx v_3^\pm \;: & \quad \frac14\mathcal O_3 + \frac34\mathcal O_8 + \frac12\mathcal O_1^\mp + \frac12\mathcal O_2^\pm + \mathcal O_3^\pm \;.
\label{eigenvaluesv3}
\end{align} \end{subequations}
In the $\mx v_{3,8}$ subspace, the operator under consideration has the following form:
\begin{equation}
	\frac14 \begin{pmatrix} 4\mathcal O_1^++4\mathcal O_1^-+\mathcal O_2^++\mathcal O_2^-+\mathcal O_3^++\mathcal O_3^- & \sqrt3(\mathcal O_2^++\mathcal O_2^--\mathcal O_3^+-\mathcal O_3^-) \\ \sqrt3(\mathcal O_2^++\mathcal O_2^--\mathcal O_3^+-\mathcal O_3^-) & 3(\mathcal O_2^++\mathcal O_2^-+\mathcal O_3^++\mathcal O_3^-) \end{pmatrix} \;.\label{eigenvaluesv38}
\end{equation}
In our case we will have that $\mathcal O_2^++\mathcal O_2^- = \mathcal O_3^++\mathcal O_3^-$, such that this part is also diagonal.

\section{Sums at finite temperature}
In this \appendixname, all integrals and sums are assumed to be part of suitably regularized multidimensional integrals, such that we do not need to care about convergence.

Consider the most general (up to a multiplicative constant) second-order polynomial $z^2+az+b$ with complex conjugate (nonreal) roots. We have that
\begin{equation}
	T \sum_{n=-\infty}^{+\infty} \ln\Big((2\pi nT)^2 + a(2\pi nT) + b\Big) = \int_{-\infty}^{+\infty} \frac{dp}{2\pi} \ln(p^2+ap+b) + T \ln(1-e^{\frac iT z_+})(1-e^{-\frac iT z_-}) \;,
\end{equation}
where $z_\pm = -\frac a2 \pm i \sqrt{b-\frac{a^2}4}$, the roots of the polynomial. In the case considered in this paper, the polynomials under consideration are of the form $(z+\alpha T)^2+\beta$. In this case we find
\begin{equation}
	T \sum_{n=-\infty}^{+\infty} \ln\Big((2\pi n+\alpha)^2T^2 + \beta\Big) = \int_{-\infty}^{+\infty} \frac{dp}{2\pi} \ln(p^2+\beta) + T \ln(1-2e^{-\sqrt\beta/T}\cos\alpha+e^{-2\sqrt\beta/T}) \;,
\end{equation}
where we performed a shift $p\to p-\alpha T$ in the integral at the right. Using the notation \eqref{inot}, we can write
\begin{equation}
	T \int\frac{d^3p}{(2\pi)^3} \sum_{n=-\infty}^{+\infty} \ln\Big((2\pi n+\alpha)^2T^2 + \vec p^2 +\beta\Big) = \int \frac{d^4p}{(2\pi)^4} \ln(p^2+\beta) + I(\beta,\alpha,T) \;.
\end{equation}

If we start from an arbitrary polynomial function $P(z)$ with two-by-two complex conjugate zeros and with the coefficient of the term with highest power equal to one, we have that
\begin{equation} \label{finitetmaster}
	T \sum_{n=-\infty}^{+\infty} \ln P(2\pi nT) = \int_{-\infty}^{+\infty} \frac{dp}{2\pi} \ln P(p) + T \ln\left(\prod_{z_0}(1-e^{\sgn(\Im(z_0))\frac iT z_0})\right) \;,
\end{equation}
where the product goes over all zeros $z_0$ of the polynomial $P(z)$, and $\sgn(\Im(z_0))$ is the sign of the imaginary part of the zero. The roots of a polynomial can be easily found numerically, making numeric evaluation straightforward.

\bibliography{biblio}

\end{document}